\documentclass[%
 reprint,
 amsmath,amssymb,
 aps,
]{revtex4-2}
\usepackage{graphicx}% Include figure files
\usepackage{dcolumn}% Align table columns on decimal point
\usepackage{bm}% bold math
\usepackage[utf8]{inputenc}

\usepackage{siunitx} %package for using Si units in formulas

\usepackage{physics}
\usepackage{amssymb}
\usepackage{amsmath}

\usepackage[dvipsnames]{xcolor}
\usepackage{cuted}
\usepackage{mathtools}
\usepackage{lineno}

\begin{document}
%\preprint{APS/123-QED}

\title{Many-body interferometry with semiconductor spins}%

\author{Daniel Jirovec$^1$}
\thanks{These authors contributed equally to this work}
\email{d.jirovec@tudelft.nl}
\author{Stefano Reale$^{1*}$}
\author{Pablo Cova-Farina$^{1}$}
\author{Christian Ventura-Meinersen$^{1}$}
\author{Minh T. P. Nguyen$^{1}$}
\author{Xin Zhang$^1$}
\author{Stefan D. Oosterhout$^{1,2}$}
\author{Giordano Scappucci$^1$}
\author{Menno Veldhorst$^1$}
\author{Maximilian Rimbach-Russ$^1$}
\author{Stefano Bosco $^1$}
\author{Lieven M. K. Vandersypen$^1$}
\email{L.M.K.Vandersypen@tudelft.nl}

\affiliation{
$^1$QuTech and Kavli Institute of Nanoscience, Delft University of Technology, 2600 GA Delft, The Netherlands\\
$^2$Netherlands Organisation for Applied Scientific Research (TNO), 2628 CK Delft, The Netherlands\\}
\date{\today}

\begin{abstract}

Quantum simulators enable studies of many-body phenomena which are intractable with classical hardware. Spins in devices based on semiconductor quantum dots promise precise electrical control and scalability advantages, but accessing many-body phenomena has so far been restricted by challenges in nanofabrication and simultaneous control of multiple interactions. Here, we perform spectroscopy of up to eight interacting spins using a 2x4 array of gate-defined germanium quantum dots.
The spectroscopy protocol is based on Ramsey interferometry and adiabatic mapping of many-body eigenstates to single-spin eigenstates, enabling a complete energy spectrum reconstruction. As the interaction strength exceeds magnetic disorder, we observe signatures of the crossover from localization to a chaotic phase marking a step towards the observation of many-body phenomena in quantum dot systems. 

\end{abstract}
\maketitle

Interacting many-body quantum systems are at the heart of some of the most intriguing phenomena in physics, ranging from high-temperature superconductors~\cite{Keimer2015} to spin liquids~\cite{Balents2010}. However, as system size grows, they can become exponentially hard to simulate using classical computations.
Analog quantum simulation platforms, such as cold atoms in quantum gas microscopes~\cite{Bloch2012}, atomic tweezer arrays~\cite{Browaeys2020}, superconducting circuits~\cite{Roushan2017, Andersen2025}, and ion trap processors~\cite{Blatt2012, Senko2014}, have proven invaluable tools for studying such systems by mimicking many-body Hamiltonians in controlled laboratory settings, providing a clear path to insights into synthetic quantum matter. Compared to these platforms, arrays of gate-defined semiconductor quantum dots offer several advantages~\cite{Barthelemy2013}. Unlike most other platforms, they combine single-site addressability with a native implementation of the Fermi-Hubbard model, leading to the observation of many intriguing phenomena~\cite{Hensgens2017, Dehollain2020, Kiczynski2022, Wang2022, Hsiao2024}. 
Furthermore, at half-filling, the Fermi-Hubbard Hamiltonian exhibits particle-hole symmetry and reduces to an effective Heisenberg spin model, enabling quantum dots to function as high-performance spin qubits with demonstrated millisecond-scale coherence times and single-qubit gate fidelities exceeding 99\%~\cite{Stano2022, Burkard2023}. In the context of quantum simulations, the resulting Heisenberg interactions between localized spins have been used to explore interesting quantum phases at small scales~\cite{Diepen2021, Qiao2020, Wang2023, Farina2025}.

Many exotic phases leave their fingerprint in the energy spectrum, through the emergence of energy gaps or changes in the level statistics~\cite{Avishai2002, Nandkishore2015, Roushan2017}. For some systems, the energy spectrum can be obtained on practical timescales with classical methods, therefore also serving as a convenient validation tool~\cite{Senko2014}. Measuring the many-body energy spectrum is, however, not trivial as traditional spectroscopic methods are often limited to low energy excitations and are subject to selection rules that further narrow their application window.

Here, we leverage the strengths of spin qubits by utilizing a 2x4 array of hole spin qubits in Ge/SiGe to implement many-body Ramsey interferometry~\cite{Roberts2024}. This protocol extends the well-known Ramsey interferometry, used to characterize single-qubit energy-level splittings, to many-body states by adiabatically turning on and off interactions~\cite{Johnson2011, Veldhorst2015, Zajac2018}. With this approach, illustrated in Fig.~\ref{fig:Device_Method}A, we initialize and disentangle superposition states and perform readout in the isolated, single-qubit (computational) basis. By adiabatically connecting the isolated spin states to interacting many-body states, the relevant phase accumulation of the Ramsey protocol occurs in the many-body sector (Fig.~\ref{fig:Device_Method}A-C). The acquired phase is therefore a direct measure of the many-body energy level spacing (Fig.~\ref{fig:Device_Method}D). 

We apply this method to the reconstruction of the energy spectrum of extended spin chains and showcase access to higher excited states. From the extracted energy levels, we infer global system properties sheding light on the transition from local observables to information about extended many-body states.

\section*{Experimental device and model}

\begin{figure}
    \centering
    \includegraphics[width=0.45\textwidth]{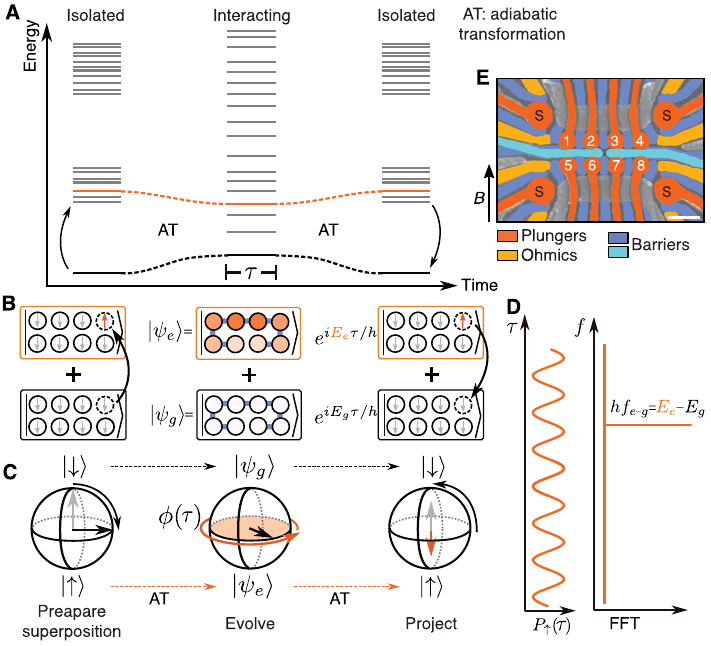}
    \caption{\textbf{Many-body Ramsey interferometry protocol.}  
    (\textbf{A}) Eigenenergies of the Heisenberg Hamiltonian with on-site disorder as a function of time. The eigenstates in the isolated and interacting regimes are connected by adiabatic transformations (AT). Waiting for a time $\tau$ in the interacting regime allows for the superposition states to acquire a relative phase. 
    (\textbf{B}) In the isolated regime, one spin can be initialized in a superposition state of $\uparrow$ and $\downarrow$. When interactions are turned on, $\ket{\downarrow}$ ($\ket{\uparrow}$) is adiabatically transformed to $\ket{\psi_g}$ ($\ket{\psi_e}$) and the initial spin excitation can delocalize over the array: the orange shades encode the probability of finding the spin excitation on a specific site. After a time $\tau$ and returning to the isolated basis, the states will have acquired a phase $E_g\tau/h$ and $E_e\tau/h$, respectively. 
    (\textbf{C}) Bloch-sphere representation of the many-body Ramsey protocol. The relevant phase accumulation occurs in the interacting regime. At the end, the accumulated phase is projected back onto the readout basis.
    (\textbf{D}) The $\uparrow$-probability of the qubit initialized in a superposition will oscillate with a frequency $hf = E_e-E_g$ allowing extraction of the many-body energy level spacings.
    (\textbf{E}) False-colored scanning electron microscope image of a nominally identical device to the one measured. Four charge sensors (S) in the corners of the device are used for RF-reflectometry charge sensing. The plunger gates (orange) $p_i$ and the barrier gates (blue and teal) $b_{ij}$ confine eight unpaired charges in a 2x4 arrangement of quantum dots. An external magnetic field $B = \SI{10}{\milli\tesla}$ is applied as indicated. The scale bar is 200~nm. 
    }
    \label{fig:Device_Method}
\end{figure}

The device is fabricated on a Ge/SiGe heterostructure with a quantum well buried 50 nm beneath the surface hosting a two-dimensional hole gas. Once cooled in a dilution refrigerator, voltages applied to metallic electrodes (gates) define the potential landscape in which we trap eight unpaired hole spins in a 2x4 arrangement of quantum dots (Fig. \ref{fig:Device_Method}E and Supplementary  section~\ref{sec:Setup}).

With a single unpaired spin in each quantum dot, the system is well described by a Heisenberg Hamiltonian with on-site disorder and a spin-orbit-induced spin-flip term~\cite{Zhang2024, SaezMollejo2025, Nguyen2025, Chou2025}:
\begin{equation}
    H = \frac{1}{2}\sum_{i}(E_{Z,i} \sigma_{i,z}+\Delta_{SO,i}\sigma_{i,x})+\frac{1}{4}\sum_{<i,j>} J_{ij}~\vec{\sigma}_i \cdot R_z (2 \gamma_{ij}) \vec{\sigma}_{j}
\label{eq:4spinHam}
\end{equation}
\noindent where $E_{Z,i} = g_i \mu_B B$ is the Zeeman energy and $ \Delta_{SO, i}$ encodes the spin-orbit induced avoided crossing. Here, $B = \SI{10}{\milli\tesla}$ is the external in-plane magnetic field, $\mu_B$ the Bohr magneton, $\hbar$ the reduced Planck constant, $g_i$ the effective g-factor of the spin in dot $i$, $\vec{\sigma}=(\sigma_x, \sigma_y, \sigma_z)$ the vector of Pauli matrices, and $J_{ij}$ the nearest-neighbor exchange interaction. The term $\Delta_{SO,i}$ originates from the different orientations of the principal in-plane g-tensor components, which can be substantial for holes in planar Ge~\cite{Hendrickx2024}, as well as intrinsic spin-orbit contributions~\cite{Geyer2024, SaezMollejo2025}. To capture the resulting anisotropy in the exchange interaction,  $R_z (2 \gamma_{ij})$ denotes a rotation about the $z$-axis by an angle $2 \gamma_{ij}$~\cite{Nguyen2025} (Supplementary section~\ref{sec:Model}). We note that equivalent formulations of eq.~\ref{eq:4spinHam} utilize isotropic exchange but explicitly introduce quantization axis tilts through site dependent g-tensors~\cite{Qvist2022, Hendrickx2024, Wang2024}.    

We can develop intuition for the resulting system dynamics by using the magnon quasiparticle representation, where the exchange interactions $J_{ij}$ constitute nearest-neighbor hopping amplitudes that enable coherent spin excitation transport across the lattice. The locally different g-factors introduce disorder through site-dependent Zeeman energies $\Delta E_{Z,ij} = (g_i-g_j)\mu_BB=\Delta g_{ij}\mu_BB $, acting as random on-site potentials that compete with the coherent hopping dynamics~\cite{Farina2025}. We find effective g-factors in the range 0.3-0.47 such that $\Delta E_{Z,ij} \ll E_{Z,i}, E_{Z,j}$ for all adjacent pairs $ij$ (see below).

In the regime where the exchange coupling strengths satisfy $J_{ij}\ll \Delta E_{Z,ij} \ll E_{Z,i}, E_{Z,j} $, spin excitations remain localized to their initialization sites due to the dominant single-particle Zeeman energies. The Hamiltonian eigenstates organize into well-separated spin manifolds characterized by z-component of total angular momentum, so the total amount of $\ket{\uparrow}$ sites, with the $n$-th manifold spanning all states containing exactly $n$ $\ket{\uparrow}$ spins (see Fig.~\ref{fig:Device_Method}A, B). Within this localized regime, energy level correlations are absent~\cite{Nandkishore2015, Roushan2017} since the large Zeeman energy differences suppress inter-site quantum tunneling, rendering each spin effectively isolated. As the exchange coupling $J_{ij}$ increases relative to the Zeeman disorder $\Delta E_{Z,ij}$, quantum tunneling enables spin excitation delocalization across the array, generating coherent spin-wave eigenmodes~\cite{Jurcevic2015} and correlations among the energy levels~\cite{Roushan2017}. We also note that, in the strongly interacting regime where $J_{ij}$ approaches the characteristic Zeeman energies $E_{Z,i}, E_{Z,j}$, the $\Delta_{SO}$ term induces hybridization between distinct spin manifolds, breaking the spin conservation of isotropic Heisenberg models. The strong spin-orbit coupling of holes in Ge, therefore, enables the study of unconventional and potentially non-trivial phenomena still largely unexplored by theory~\cite{Chou2025}.

\section*{Interferometry of non-interacting spins and two-spin chains}
\begin{figure*}
    \centering
    \includegraphics[width=0.99\textwidth]{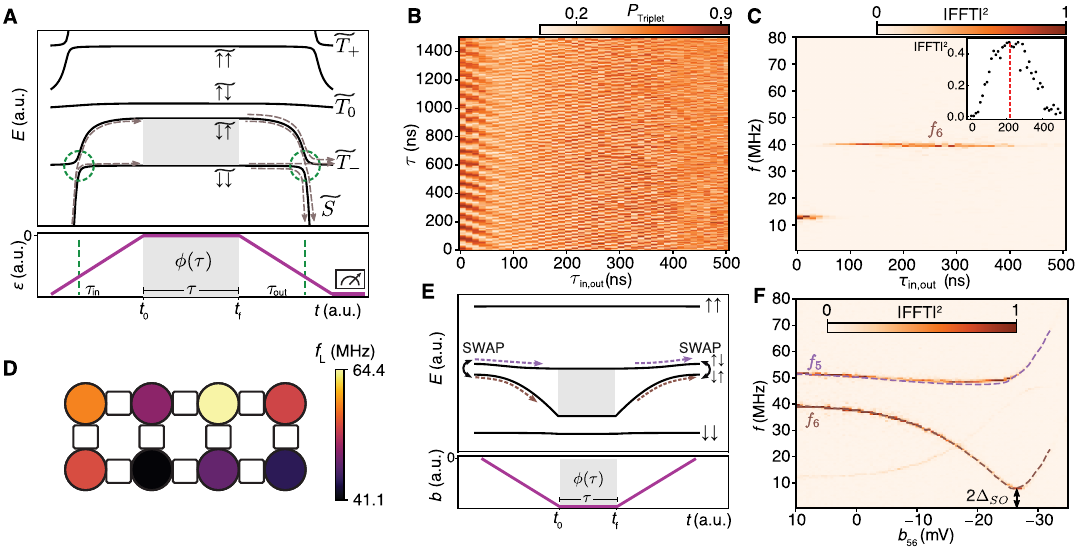}
    \caption{\textbf{Interferometry implementation.} (\textbf{A}) Top: Energy level diagram of a two-spin system as a function of the lab time. Bottom: the corresponding pulse on the detuning $\epsilon$. The grey dashed lines indicate the state trajectories resulting from the ramp in detuning $\epsilon$, passing through the spin-orbit induced avoided crossing (green dashed circles). After letting the system acquire relative phase $\phi$ for a time $\tau = t_f-t_0$, another ramp back towards the readout point closes the interference loop of the two trajectories. We can read the resulting triplet probability, which will oscillate at a frequency $hf = h\phi(\tau)/\tau=E_{\downarrow\uparrow}-E_{\downarrow\downarrow}$. Any phase accumulated before $t_0$ and after $t_f$ only adds a constant offset to $\phi$. 
    (\textbf{B}) Probability of finding the final state in a triplet as a function of ramptime $\tau_{ramp}$ and free evolution time $\tau$ for spin pair 5-6. Two sets of oscillations can be seen which are further evidenced by the fast Fourier transform (FFT) in C.
    (\textbf{C}) Squared norm of FFT of B. The inset shows the FFT amplitude at $f_6$ as a function of $\tau_{in,out}$. The red dashed line marks the ramp time corresponding to $P_{LZ} = 0.5$.
    (\textbf{D}) Extracted Larmor frequencies $f_L$ defining the magnetic disorder landscape.  
    (\textbf{E}) Energy spectrum as a function of time and barrier voltage (as sketched in the bottom panel). After the initialization through the avoided crossing we can choose to apply a local SWAP operation. If no SWAP (a SWAP) is applied, we follow the lower (higher) branch. After adiabatically activating the exchange interaction, we let the system acquire a relative phase. The adiabatic turn-off is followed by no (a) SWAP in case we selected the lower (higher) branch. Subsequently, we close the interference loop like in A.
    (\textbf{F}) Sum of FFTs of oscillations as a function of the inter-qubit barrier $b_{56}$, using 200 ns ramps. When turning interactions on and off without SWAPs applied, we observe the lower frequency. In this case, spin 6 is prepared in a superposition and we label the frequency $f_6$. When we apply a SWAP before and after the phase acquisition and adiabatic exchange turn-on and off, we observe the upper frequency. Here spin 5 is initialized in a superposition and we call this frequency $f_5$. The brown and purple dashed lines are the fitted model (Eq.~\ref{eq:4spinHam}) from which we extract the exchange interaction strength at each point. The minimum of $f_6$ corresponds to $2\Delta_{SO}$.}
    \label{fig:BIMBS}
\end{figure*}

As a starting point, we demonstrate interferometry of non-interacting spins. Standard spin-qubit Ramsey protocols rely on microwave-driven single-spin qubit rotations; however, the small Zeeman splitting at low magnetic fields, typical of our experiments, renders such resonant driving impractically slow compared to relevant decoherence timescales in our device~\cite{Bosco2021,AbadilloUriel2023}.
For this reason, we initialize superpositions of spin states by leveraging the spin-orbit induced avoided crossing~\cite{Petta2010, Ivakhnenko2023}. In Fig.~2A we plot the energy levels of a two-spin system as a function of time when pulsing the energy detuning between adjacent dots $\epsilon$ as shown. The approximate eigenstates, symbolized by $\sim$, are also labeled. In the following we omit the $\sim$, implicitly referring to the approximate Hamiltonian eigenstates. At $t=0$, we initialize $|S\rangle$ at $|\epsilon|\gg 0$, where both charges reside in the same dot ((2,0) or (0,2) charge configuration). We then ramp $\epsilon$ to zero, passing semi-adiabatically through the avoided crossing (dashed green circles)~\cite{Kelly2025}. This Landau-Zener passage initializes a superposition state $\ket{\psi} = P_{LZ}\ket{S}+(1-P_{LZ})\ket{T_-}$, with $P_{LZ}= \exp^{-2\pi\Delta_{SO}^2/\hbar v}$ being the Landau-Zener probability~\cite{Zener1932, Landau1932, Ivakhnenko2023}. $v = \frac{dE}{d\tau_{in}}$ is the ramp speed, $E$ is the energy difference of the states involved in the transition and $\tau_{in}$ is the ramp-in time. After the avoided crossing, the detuning-ramp transforms $\ket{S}$ into $\ket{downarrow\uparrow}$ adiabatically (with respect to $\Delta E_Z$), which effectively initializes the state $\ket{\psi} = \ket{\downarrow}\otimes (P_{LZ}\ket{\uparrow}+(1-P_{LZ})\ket{\downarrow})$, i.e. only one of the spins in the pair is in a superposition.  We then let the system acquire a relative phase $\phi(\tau)=\frac{(E_{\downarrow\uparrow}-E_{\downarrow\downarrow})\tau}{\hbar}$. Another semi-adiabatic passage through the avoided crossing, with a ramp time $\tau_{out} = \tau_{in}$ rotates the state back into the readout basis, closing the interference loop. Finally, we read out the resulting triplet probability (Supplementary section~\ref{sec:Setup} and \ref{sec:Initialization_pulses}). By choosing $\tau_{in,out}$ such that $P_{LZ} = 0.5$, this sequence is analogous to a traditional Ramsey experiment, and the oscillations we observe are known as Landau-Zener-Stückelberg-Majorana oscillations~\cite{Ivakhnenko2023}.

Fig.~\ref{fig:BIMBS}B plots the probability of measuring a triplet state after an experiment as described in Fig.~\ref{fig:BIMBS}A, as a function of $\tau_{in,out}$ and waiting time in (1,1) $\tau$ for the spin pair 5-6. Fig.~\ref{fig:BIMBS}C plots the squared norm of the fast Fourier transform (FFT) of Fig.~\ref{fig:BIMBS}B where we can discern two main frequency components. For short ramp times we observe oscillations with a frequency set by the Zeeman energy difference between the two spins: $hf_{5-6} = \Delta g \mu_B B$, known as $S-T_0$ oscillations. At longer ramp times, the low-frequency component disappears indicating adiabaticity with respect to $\Delta E_z$. However, we see a higher-frequency component appearing with a maximum FFT amplitude around 220~ns (see inset of Fig.~\ref{fig:BIMBS}C), corresponding to the energy difference of the states $\ket{\downarrow_5\downarrow_6}\leftrightarrow\ket{\downarrow_5\uparrow_6}$. The maximum in visibility corresponds to $P_{LZ}\approx 0.5$ and in the localized regime the oscillation frequency $f_6$ corresponds to the Larmor frequency of spin 6, $f_6 = g_6\mu_B B/h$, in this case (Supplementary section~\ref{sec:Larmor_freqs} on how to determine the correspondence between dot position and measured frequency). For $\tau_{in,out} > \SI{400}{\nano\second}$, also this frequency component tends to fade away, which indicates adiabaticity, now with respect to $\Delta_{SO}$.  

To extract the other Larmor frequency in the pair, namely spin $5$, we can perform a local SWAP before and after the relative phase accumulation (Supplementarysection~\ref{sec:SWAP_calibration}). 
Fig.~\ref{fig:BIMBS}D summarizes the extracted Larmor frequencies for all the spins in the array which determines the on-site magnetic disorder landscape for the system Hamiltonian.

For the many-body interferometry protocol, we adiabatically turn on exchange and reconstruct the two-qubit energy spectrum with the pulse scheme depicted in Fig.~\ref{fig:BIMBS}E. Similar to Fig.~\ref{fig:BIMBS}A, we schematically plot the energy spectrum as a function of time, but now only after the initialization and before the readout part. Here, only the barrier voltage $b$ is varied, maintaining $\epsilon=0$. If we do (do not) apply a SWAP prior to and after the phase accumulation, we follow the upper (lower) energy branch. 
Fig.~\ref{fig:BIMBS}F displays the sum of the FFTs of two experiments, one with and one without SWAPs. Without (with) SWAPs we record $f_6$ ($f_5)$ as a function of $b_{56}$. For each barrier value, we record oscillations for up to $\SI{1}{\micro\second}$ and extract the highest peak position from the FFT through a Gaussian fit. We then fit the extracted frequencies to the full Hamiltonian (Eq.~\ref{eq:4spinHam}) for two spins to obtain $J_{56}$ for each $b_{56}$ value. The size of the avoided crossing can be directly measured from the minimum frequency of $f_6$. As we do not have access to different magnetic field orientations, we are not able to fully determine the angle $\gamma_{56}$. However, with the constraint of a finite minimum $f_6$, here and in the following, we take $\gamma_{ij} = \pi/2$. The dashed lines overlaid in Fig.~\ref{fig:BIMBS}F show $f_5$ and $f_6$ using the extracted $J_{56}$ for each $b_{56}$. The additional, faintly visible low-frequency line corresponds to the energy difference between $f_5$ and $f_6$ (hence to the $S$-$T_0$ oscillation frequency) and it can be attributed to an imperfect SWAP (Supplementary section~\ref{sec:Adiabatic_ramps}).

We repeat the experiments in Fig.~\ref{fig:BIMBS} for most qubit pairs and extract the calibrated initialization and readout ramp times, SWAP times, and exchange values as a function of gate voltage (Supplementary section~\ref{sec:Larmor_freqs}).

\section*{Interferometry of extended spin chains}

\begin{figure}[t]
    \centering
    \includegraphics[width=0.5\textwidth]{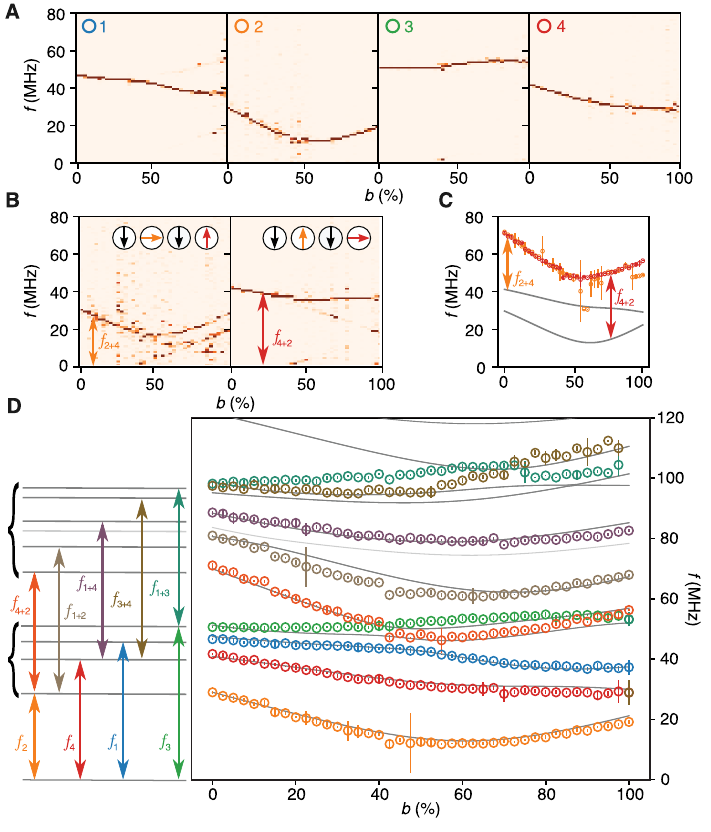}
    \caption{\textbf{Interferometry of extended spin chains.} (\textbf{A}) FFTs of oscillations for spins in a four-spin chain as a function of global barrier gate voltage, in percent, where $b = 0\%$ corresponds to $J\ll \Delta E_Z$ and $b = 100\%$ to $J> \Delta E_Z$.  The panels are ordered to reflect the position  of the spin initialized in a superposition in the array. The four frequencies $f_i$ allow  reconstruction of the first spin manifold.
    (\textbf{B}) FFTs of oscillations for the initial states in the inset. On the left (right), we initialize spin 2 (4) in a superposition, while spin 4 (2) is initialized in $\ket{\uparrow}$. 
    (\textbf{C}) To reconstruct the energy levels of the second manifold we add $f_4$ ($f_2$) measured in A, represented by the solid gray lines, to the corresponding frequencies measured in B. We see that the data points lie almost perfectly on top of each other, and return the frequency of the state $\ket{\downarrow\uparrow\downarrow\uparrow}$ from two complementary measurements.
    (\textbf{D}) Reconstructed spectrum of the four-spin chain 1-2-3-4. When redundant frequencies have been measured, we chose the one with the fewest artifacts. The gray lines show the spectrum obtained from the fitted model.}
    \label{fig:4spin_spectroscopy}
\end{figure}

The adiabatic turn-on of exchange can be simultaneously applied and uniformly implemented along extended spin chains by appropriately varying all the corresponding barrier gate voltages. To reconstruct the energy spectrum, we repeat the same interferometry experiment with all the different combinations of initial states. Mapping out the first spin manifold, containing a single spin excitation, we prepare spin $i$ in a superposition state $\ket{\rightarrow} = (\ket{\uparrow}+\ket{\downarrow})/\sqrt{2}$ and measure $f_i$ as a function of exchange. For the second spin manifold, we prepare spin $j$ in $\ket{\uparrow}$ and spin $i$ in a superposition. The total frequency will then be the measured frequency $f_i$ summed with the previously obtained $f_j$ from the lower manifold. Similar arguments hold for higher spin manifolds. We adopt the convention that in $f_{i+j,k,l}$ the first index corresponds to the spin initialized in a superposition, while the other indeces correspond to the spins that are $\ket{\uparrow}$. 

As an example for a chain of four spins, we plot the extracted frequencies as a function of global barrier voltage of a chain formed by spins 1-2-3-4 for the first manifold in Fig.~\ref{fig:4spin_spectroscopy}A. In Fig.~\ref{fig:4spin_spectroscopy}B we showcase measurements to reach the second manifold. In the left panel, we initialize the state $\ket{\downarrow_1 \rightarrow_2 \downarrow_3 \uparrow_4}$ measuring $f_{2+4}$. In the right panel, we instead initialize $\ket{\downarrow_1 \uparrow_2 \downarrow_3 \rightarrow_4}$ and measure $f_{4+2}$. Fig.~\ref{fig:4spin_spectroscopy}C plots $f_{2+4}$ (orange) and $f_{4+2}$ (red) after adding the corresponding frequencies measured for the first spin manifold, represented by the solid gray lines. We clearly see that the effective frequency we extract for the two initial states is the same, allowing us to determine the actual frequency corresponding to the state $\ket{\downarrow_1 \uparrow_2 \downarrow_3 \uparrow_4}$ in two ways. This redundancy can become a useful verification tool in the energy level reconstruction.

In Fig.~\ref{fig:4spin_spectroscopy}D we present the experimentally reconstructed energy spectra for both the first and second spin manifolds of this four-spin chain. The measured transition frequencies (circles) are color-coded according to the double arrows in the schematic energy level diagram (left panel). The gray lines are the fitted model (Eq.~\ref{eq:4spinHam}) to the data. The values for the $g$-factors, $\Delta_{SO}$ for each pair and the functional form of $J_{ij}$ versus $b_{ij}$ are close to those extracted from the data on pairs of spins (Supplementarysection~\ref{sec:Exchange_extraction} for further details). One spectral transition (light gray) remains experimentally unresolved due to insufficient SWAP gate fidelity.

\section*{Inferring global system properties}
\begin{figure*}
    \centering
    \includegraphics[width=0.95\linewidth]{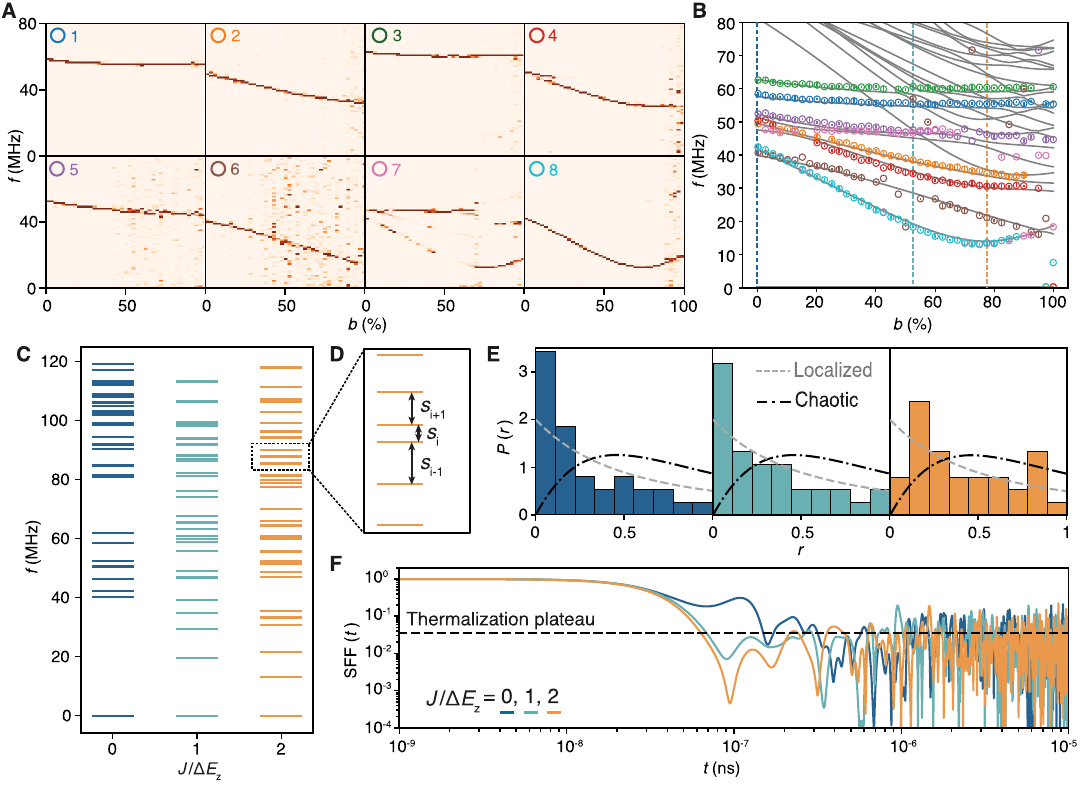}
    \caption{\textbf{Inferring global system parameters.}
    (\textbf{A}) FFTs of oscillations for a chain of eight spins. The panels are ordered to reflect the position of the measured spin in the array. The barrier gates are all swept together from low exchange $J\ll \Delta E_Z$ ($b = 0\%$) to high exchange $J>\Delta E_z$ ($b = 100\%$). Except for spin 7, we can always clearly identify the main frequency.
    (\textbf{B}) Extracted frequencies from A with overlaid fitted model in gray. At the colored dashed lines we reconstruct higher spin manifolds.
    (\textbf{C}) Energy levels of the first and second spin manifold extracted for three different values of $\frac{J}{\Delta E_Z}$ highlighted in B. We observe clearly separated manifolds for $\frac{J}{\Delta E_Z}\approx 0$. For $\frac{J}{\Delta E_Z}\approx 2$ the energy levels tend to repel each other signaling correlations and, consequently, a tendency for chaotic behavior.
    (\textbf{D}) Visualization of the definition of $s_i$.
    (\textbf{E}) Energy level uniformity distribution for the three exchange configurations calculated from the energy levels extracted in C. The gray  dashed (black dashed dotted) line represents the expected theoretical distribution for localized (chaotic) systems. The data displays a trend to more chaotic behavior as interactions increase, signaled by $P(r\to0)\to 0$ for $\frac{J}{\Delta E_Z}\approx 2$. 
    (\textbf{F}) Spectral form factor calculated for the energy levels extracted in C. We observe a systematic trend to reach a deeper early-time minimum smaller than the thermalization plateau (black dashed line at a value of $1/N$) as we increase the interaction strength. For longer times, oscillatory behavior masks the universal features of the SFF, since we do not average over disorder realizations.
    }
    \label{fig:SFF}
\end{figure*}

Adding all the spins in the array we perform comprehensive interferometry of an eight-spin chain (dots 6-5-1-2-3-4-8-7) as a function of global exchange coupling, reconstructing the energies of the first spin manifold (Fig.~\ref{fig:SFF}A). As for the shorter spin chains, we fit Eq.~\ref{eq:4spinHam} to the spectrum in Fig.~\ref{fig:SFF}B, once again finding good agreement after only small modifications to the fit parameters extracted from pairwise characterization, demonstrating robust control of seven simultaneously activated exchange interactions.

At interaction strengths $J/\Delta{E_z}$ $\approx$ 0, 1, and 2 (vertical dashed lines in Fig.~\ref{fig:SFF}B), we extend spectroscopic reconstruction to the second manifold (Fig.~\ref{fig:SFF}C) and implement statistical analysis of the energy levels to reveal signatures of quantum chaos. 
Higher excitation manifolds can in principle be reconstructed as well, through sequential SWAP protocols or driving, at the cost of added complexity in the pulse sequence.

From the energy levels $\{E_i\}$ we can compute the energy level uniformity $r_i = \frac{\min\{s_i, s_{i+1}\}}{\max \{s_i, s_{i+1}\}}$, where $s_i = E_i-E_{i-1}$~\cite{Roushan2017} (see Fig.~\ref{fig:SFF}D). The level uniformity can capture energy level repulsion, characteristic of quantum chaotic systems, as the probability distribution $P(r_i\to 0) \to 0$. Such a correlation of energy level spacings is absent in localized systems, leading to randomly distributed energy levels. Fig.~\ref{fig:SFF}E displays $P(r_i)$ as we increase the interaction strength. Indeed we observe a reduction of counts for $P(r_i\to 0)$, signaling increased correlations in the energy levels. The gray dashed (black dashed dotted) lines show the ideal distributions for localized (chaotic) systems, obtainable in large ensembles, as further detailed in  Supplementary section~\ref{sec:Model}.

Additionally, from $\{E_i\}$, we compute the spectral form factor, defined as $\text{SFF}(t)\equiv \frac{1}{N^2}\sum_{i,j}e^{i(E_i-E_j)t/\hbar}$, where $N$ is the system size. This quantity can capture time correlations in the energy levels beyond nearest neighbor, making it a potentially more sensitive probe of quantum chaotic systems.  Similar to $r$, also the SFF displays universal signatures that are different between localized and chaotic systems, described in more detail in Supplementary section~\ref{sec:Model}. At relatively short times, the SFF develops a dip with a minimum value $\approx\frac{1}{N}$ ($\approx\frac{1}{N^{3/2}}$) for localized (chaotic) systems. Subsequently, energy level correlations lead to a linear in $t$ ramp, up to the thermalization limit $\frac{1}{N}$. Since the thermalization limit corresponds to the dip in the localized case, no linear ramp will develop, as correlations are absent~\cite{MatsoukasRoubeas2024}. Importantly, the SFF requires sufficient statistics to display all the universal features, which are otherwise masked by noise. However, in Supplementary section~\ref{sec:Model} we show that the early time behavior is robust even without averaging. Indeed, the SFF displayed in Fig.~\ref{fig:SFF}E systematically develops a deeper early time dip around $t~\approx10^{-7}$. The systematic trends of both $P(r_i)$ and the SFF as a function of interaction strength suggest a transition from localization to quantum chaotic behavior in the eight-spin chain. This result showcases how adiabatic transformations enable inferring global system parameters from local observables in quantum dot based spin qubit systems.

\section*{Conclusion}
In summary, we have demonstrated many-body interferometry of spin chains with up to eight spins. To perform the interferometry we leveraged a spin-orbit induced avoided crossing, which allowed us to initialize and disentangle superposition states using only on baseband pulses. Furthermore, to access additional initial states, we implemented local SWAPs. By initializing different states and adiabatically turning on exchange interactions, we probed the energy levels in two-, four-, and eight-spin chains supporting our findings with extensive simulations.

After mapping out the energies of higher spin manifolds, we 
computed the energy level uniformity and the spectral form factor as a function of interaction strength, finding trends consistent with a crossover from localized to quantum chaotic behavior.
Further work will be required to unveil more universal features in the spectral form factor as well as the energy level uniformity. 

We note that, as the energy spectra become more dense and more avoided crossings appear, tailored transformation strategies need to be implemented to maintain adiabaticity for various initial states, an active area of research in the context of quantum annealing~\cite{Hauke2020}. The present method combined with our earlier studies of magnon propagation~\cite{Farina2025}, will allow us to engineer multi-spin Hamiltonians and probe their dynamics to reveal signatures of many-body localization.

\subsection*{Acknowledgements}
We thank members of the Vandersypen, Veldhorst, Scappucci, Rimbach-Russ and Bosco groups for stimulating discussions. We acknowledge technical support by O. Benningshof, J. D. Mensingh, T. Orton, R. Schouten, R. Vermeulen, R. Birnholtz, E. Van der Wiel, B. Otto, D. Brinkman and S. de Snoo.\\
This work was financed by the Advanced Grant QuDoFH (882848) and the ERC Starting Grant QUIST (850641) from the European Research Council (ERC) under the European Union’s Horizon 2020 research and innovation programme, the NWO Spinoza prize awarded to L.M.K.V. by the Netherlands Organisation for Scientific Research (NWO/OCW), and by NWO/OCW as part of Quantum Limits (project number SUMMIT.1.1016)
This work is partly supported by the Army Research Office under Award Number: W911NF-23-1-0110. The views and conclusions contained in this document are those of the authors and should not be interpreted as representing the official policies, either expressed or implied, of the Army Research Office or the U.S. Government. The U.S. Government is authorized to reproduce and distribute reprints for Government purposes notwithstanding any copyright notation herein.
This research was sponsored in part by The Netherlands Ministry of Defence under Awards No. QuBits R23/009. The views, conclusions, and recommendations contained in this document are those of the authors and are not necessarily endorsed nor should they be interpreted as representing the official policies, either expressed or implied, of The Netherlands Ministry of Defence. The Netherlands Ministry of Defence is authorized to reproduce and distribute reprints for Government purposes notwithstanding any copyright notation herein.

\subsection*{Data availability statement}
The data, analysis code, and simulation code will be made available in a Zenodo repository (10.5281/zenodo.17457400).

\subsection*{Competing interests:}
M.V., G.S., and L.M.K.V. are Founding Advisors of Groove Quantum BV and declare equity interests. The remaining authors declare that they have no competing interests.

\bibliography{Bibliography}
\clearpage

\widetext
\begin{center}
\textbf{\large Supplemental Materials: Manybody interferometry with spins in Germanium}
\end{center}

%%%%%%%%%% Merge with supplemental materials %%%%%%%%%%
%%%%%%%%%% Prefix a "S" to all equations, figures, tables and reset the counter %%%%%%%%%%
\setcounter{figure}{0}
\setcounter{section}{0}
\setcounter{page}{1}
\makeatletter
\renewcommand{\thefigure}{S\arabic{figure}}
\renewcommand{\thesection}{\arabic{section}}

\section{8-spin chain Hamiltonian}
\label{sec:Model}
We model a linear chain of QDs in a Ge/SiGe heterostructure using a generalized Fermi-Hubbard Hamiltonian in the qubit frame \cite{SaezMollejo2025,Nguyen2025,Geyer2024}
\begin{equation}
     \label{eq: Fermi-Hubbard Hamiltonian}
    H_{\rm FH}  = \sum_{i,\sigma} \epsilon_i \hat{n}_{i,\sigma}-\sum_{i\neq j} t_{ij}  \textbf{c}_{i}^{\dagger} \hat{S}_{\rm rot}^{ij} \textbf{c}_j + \sum_{i} U_i  \hat{n}_{i,\uparrow } \hat{n}_{i,\downarrow}  + \sum_{i}E_{Z,i}(\hat{n}_{i,\uparrow} - \hat{n}_{i,\downarrow})   ,
\end{equation}
where the transmission matrix $\hat{S}^{ij}_{\rm rot}$ is given by
\begin{equation}
    \hat{S}_{\rm rot}^{ij}  = \exp\Big[- i \gamma_{ij}(\vec{n}_{\rm so, ij}\cdot \vec{\sigma}) \Big].
\end{equation}
Here, $\textbf{c}_i = [\hat{c}_{i,\uparrow}, \hat{c}_{i,\downarrow}]^{T}$ is the spinor annihilation operator for dot $Q_i$, $\hat{n}_{i,\sigma} = \hat{c}_{i,\sigma}^{\dagger} \hat{c}_{i,\sigma}$ is the number of operator at site $i$ with spin index $\sigma$, and $\vec{\sigma}$ is the vector of Pauli matrices. Each dot $Q_i$ is characterized by an energy $\epsilon_i$, a Zeeman splitting $E_{Z,i}$, and the charging energy $U_{i}$. The tunneling amplitude between dots $Q_i$ and $Q_j$ is $t_{ij}$. The matrix $\hat{S}_{\rm rot}^{ij}$ describes the spin tunneling events in which the tunneling particle is rotated about the axis $\vec{n}_{\rm so,ij}$ by an angle $\gamma_{ij}$.

While the Hamiltonian $H_{\rm FH}$ is general, it can be significantly simplified by using the geometry of the system. This simplification proceeds in two steps. First, as shown in Refs.~\cite{Nguyen2025, Chou2025, Shekhtman1992, Derzhko1994, Oshikawa1997}, for systems arranged in a linear chain or a closed loop, one can apply a local gauge transformation to each quantum dot such that all spin-orbit vectors  $\vec{n}_{\rm so,ij}$ align along the $Z$-direction. This transformation modifies the Zeeman terms, introducing components along the $X$ and $Y$ directions. In the second step, additional local gauge transformations—specifically, rotations about the local $Z$-axis can be applied to further simplify the Zeeman interaction, reducing it to contain only Pauli-$Z$ and Pauli-$X$ terms. As a result, the effective qubit Hamiltonian can be brought into the following simplified form:
\begin{equation}
    \label{eq: effective Hamiltonian}
    H_{\rm eff} = \frac{1}{2}\sum_{i} (E_{Z,i} \sigma_{z,i} + \omega_{i,x} \sigma_{i,x}) + \frac{1}{4}  \sum_{\langle ij \rangle} J_{ij} ~\vec{\sigma}_i \cdot R_z (2 \gamma_{ij}) \vec{\sigma}_{j} ~. 
\end{equation}
where $R_z (2 \gamma_{ij})$ denotes a rotation about the $z$-axis by angle $2 \gamma_{ij}$. In this representation, the anisotropy is distributed across both single-qubit terms and the exchange interactions. However, for a linear topology, it is always possible to transfer all anisotropy into either the single-qubit terms using another set of local gauge transformation
\begin{equation}
    \label{eq: effective Hamiltonian form 2}
    H_{\rm eff}^{'} = \sum_{i} \frac{1}{2}(\omega_{i,x}^{'} \sigma_{x,i}+\omega_{i,y}^{'} \sigma_{y,i}+\omega_{i,z}^{'} \sigma_{z,i}) + \frac{1}{4} \sum_{\langle ij \rangle} 
    J_{ij} \vec{\sigma}_i \cdot \vec{\sigma}_j \quad , 
\end{equation}
or into the exchange interaction, 
\begin{equation}
    \label{eq: effective Hamiltonian form 3}
    \tilde H_{\rm eff} = \sum_{i} \frac{1}{2} \tilde \omega_{i,z} \sigma_{z,i} + \frac{1}{4}  \sum_{\langle ij \rangle} J_{ij} ~\vec{\sigma}_i \cdot R_{y}^{T}(\theta_i) R_{z} (2 \gamma_{ij}) R_{y}(\theta_j)\vec{\sigma}_{j} ~. 
\end{equation}
where the set $\{ R_{y}^{T}(\theta_i) R_{z} (2 \gamma_{ij}) R_{y}(\theta_j) \}$ forms general rotation matrices, and the angles $\{ \theta_i \}$ are implicitly determined by $\{ \omega_{i,z}, \omega_{i,x}\}$.

The effective Hamiltonian we derive encompasses the fitted model used in Ref.~\cite{Zhang2024}. In our experimental setting, we find that the following minimal Hamiltonian provides an excellent fit to the observed low-energy spectrum
\begin{equation}
    \label{eq: fit Hamiltonian}
    H_{\text{fit}} = \frac{1}{2} \sum_{i} (\omega_{z,i} \sigma_{z,i} + \omega_{x,i} \sigma_{x,i}) + \frac{1}{4} \sum_{\langle ij \rangle} J_{ij} (-\sigma_{x,i} \sigma_{x,j} - \sigma_{y,i} \sigma_{y,j} + \sigma_{z,i} \sigma_{z,j}),
\end{equation}
with $\omega_{z,i} = E_{Z,i}$ and $\omega_{x,i} = \Delta_{SO,i}$ as shown in the main text.
We remark that the exchange interaction in Eq.~\eqref{eq: fit Hamiltonian} deviates from the standard Heisenberg exchange due to the relative sign difference between the $XX,YY$ terms and the $ZZ$ term. This structure is consistent with the theoretically predicted effective Hamiltonian $H_{\rm eff}$, and suggests that the spin-orbit angle satisfies $\gamma_{ij} \approx \frac{\pi}{2}$. Equivalently, from the perspective of Eq.~\eqref{eq: effective Hamiltonian form 2}, the system can be viewed as staggered g-tensor system, with signs of the tilted Zeeman field $\omega_{x}$ alternating between positive and negative.

We remark that the spin-orbit angle $\gamma_{ij}$ include the effect from both the (linear/cubic) spin-orbit interaction and the local g-tensor difference. It was theoretically predicted and recently shown that the native (linear/cubic) spin-orbit angle can be small compared to the fitted vale of $\pi/2$ \cite{Seidler2025}. However, it was reported in Ref.~\cite{Wang2024} that the artificial local g-tensor difference can lead to artificial spin-orbit angle approximately equal $\pi/2$. Therefore, the fitted value is consistent with theory and experiments and suggests that we can attribute most of the spin-orbit interaction in the linear chain to the local g-tensor difference. 

Nevertheless, our conclusion is not definitive, as we only have access to the low-energy spectrum of the full Hamiltonian. As a result, not all parameters can be uniquely determined. Indeed, our numerical calculations shows that a good fit can also be obtained for values of $\gamma_{ij}$ that differ significantly from $\pi/2$, provided that the transverse fields $\omega_{x,i}$ are allowed to vary. Therefore, Eq.~\eqref{eq: fit Hamiltonian} should be interpreted as a minimal model that captures the low-energy behavior of the system, rather than an exact microscopic description.

\section{Signatures of localization and quantum chaos}
\label{sec:Signature_of_loc_and_chaos}

In the following, we introduce the observables to discern localized from quantum chaotic systems and the specific form they take for the system in this work. 

\subsection{Universality class}
Having derived both the theoretical Hamiltonian $H_{\rm eff}$ and its corresponding fitted Hamiltonian $H_{\rm fit}$, we now turn to determining the random matrix universality class into which these Hamiltonians fall when the system is in the chaotic regime. Specifically, we seek to identify whether the spectral statistics align with the Gaussian Orthogonal Ensemble (GOE), Gaussian Unitary Ensemble (GUE), or Gaussian Symplectic Ensemble (GSE).

At first glance, one might expect the Hamiltonian in Eq.~\eqref{eq: fit Hamiltonian} to fall under the GUE class, as it breaks time-reversal symmetry explicitly due to the asymmetry in spin-exchange terms and the presence of Zeeman terms. However, our numerical simulations of the level statistics of system sizes up to $10$ dots consistently reveal level repulsion and spectral characteristic of the GOE ensemble.

In the following, we explain this seemingly paradoxical result. We show that despite the apparent absence of conventional time-reversal symmetry, the Hamiltonian possesses a generalized anti-unitary symmetry that ensures that the energy levels obey GOE statistics rather than GUE, consistent with our numerical observations.

The key observation is that the effective Hamiltonian $H_{\rm eff}$ can be naturally decomposed into two distinct components
\begin{equation}
    H_{\rm eff} = H_{\rm GOE} + H_{\rm GUE},
\end{equation}
where
\begin{equation}
    H_{\rm GOE} =  \frac{1}{2} \sum_{i} (\omega_{i,z} \sigma_{z,i} + \omega_{i,x} \sigma_{x,i}) + \frac{1}{4} \sum_{\langle ij \rangle} J_{ij} [ \cos(2 \gamma_{ij})  (\sigma_{x,i} \sigma_{x,j} + \sigma_{y,i} \sigma_{y,j}) + \sigma_{z,i} \sigma_{z,j}]
\end{equation}
and
\begin{equation}
    H_{\rm GUE} = \frac{1}{4} \sum_{\langle ij \rangle} J_{ij}  \sin(2 \gamma_{ij})  (\sigma_{x,i} \sigma_{y,j} - \sigma_{y,i} \sigma_{x,j}).
\end{equation}
The term $H_{\rm GOE}$ is comprised entirely of real-valued operators when expressed in the computational basis. Although $H_{\rm GOE}$ is not invariant under the conventional time-reversal operator $T_0 = i \sigma_y \mathcal{K}$, it was shown in Ref.~\cite{Avishai2002}  that it is invariant under the anti-unitary operator $T = \mathcal{K}$, where $\mathcal{K}$ denotes complex conjugation operations. This generalized time-reversal symmetry ensures that $H_{\rm GOE}$ is a real symmetric matrix, and thus its spectrum exhibits GOE statistics, even though it lacks physical time-reversal symmetry in the conventional sense.

On the other hand, $H_{\rm GUE}$ is a purely imaginary and antisymmetric operator, and thus does not commute with $T$. However, since the Zeeman terms are either much larger or comparable to the exchange energy, $||H_{\rm GOE}|| \geq H_{\rm GUE}$, leaving the effective Hamiltonian dominated by its GOE component \cite{Avishai2002}. This explains why the system consistently exhibits GOE level statistics despite the absence of explicit time-reversal symmetry in the microscopic model.

\subsection{Level statistics and spacing ratio}

In the simplest case for the GOE, we have a $2\times 2$ real-valued symmetric matrix $H$ (eigenvalues $\{\lambda_+,\lambda_-\}$) with components that are drawn from a Gaussian distribution $p(H_{ij})\propto e^{-H_{ij}^2/2}$ with eigenvalue splitting
\begin{align}
    s=\lambda_+-\lambda_-=\sqrt{(H_{11}-H_{22})^2+(2H_{12})^2}.
\end{align}
The probability distribution for this spacing is then given by
\begin{align}
    p_\text{GOE}(s)&\propto \int dH_{11}dH_{22}dH_{12} \;\delta\left(s-\sqrt{(H_{11}-H_{22})^2+(2H_{12})^2}\right)p(H_{11})p(H_{22})p(H_{12})\\
    &\propto \int_0^{2\pi}d\phi\, \int_0^\infty  dr \, r \,\delta(s-r) e^{-r^2} \propto \; s \,e^{-s^2},
\end{align}
where the proportionality is due to the appropriate normalization. 
We find that in the limit of $s\to 0$, the probability distributions approach zero, alluding to the \textit{eigenvalue repulsion phenomenon}. In Fig.~\ref{fig: matrix ensembles ideal}A we compare the eigenvalue probability distribution for the localized case ($p_{Poiss}(s)\approx e^{-s}$) to that of a chaotic system goverend by GOE statistics.  

We further introduce the level spacing ratio
\begin{equation}
r_n = \frac{\min(s_n, s_{n+1})}{\max(s_n, s_{n+1})}
\end{equation}
as it is believed and numerically confirmed across a broad range of quantum models, to serve as a robust diagnostic for quantum integrability. Here, $s_{n} = E_{n+1}-E_n$, denotes the adjacent energy level spacings of the ordered spectrum.

For integrable systems, where energy levels follow uncorrelated statistics, the distribution $P(r)$ of the spacing ratio can be derived from the Poisson ensemble and takes the form
\begin{equation}
    \label{eq: Poisson Pr distribution}
    P_{\rm Poiss}(r) = \frac{1}{(1+r)^2}.
\end{equation}
In contrast, for quantum chaotic systems in the GOE class, $P(r)$ follows a Wigner-like distribution well-approximated by the surmise
\begin{equation}
    \label{eq: GOE Pr distribution}
    P_{\rm GOE}(r) = \frac{27}{4}\frac{(r+r^2)}{(1+r+r^2)^{5/2}}. 
\end{equation} 
This makes $P(r)$ a powerful and simple tool for distinguishing between integrable and chaotic regimes, even in finite-size quantum systems assuming access to multiple realizations of the system \cite{Roberts2024}. Fig.~\ref{fig: matrix ensembles ideal}B compares the ideal spacing ration $r_n$ for integrable and chaotic systems. A distinguishing factor is that $P(r_n\to0)\to 2$ for localized systems, while $P(r_n\to0)\to 0$ in the chaotic regime, similar to what we observe in the data in the inset of Fig.~\ref{fig:SFF}D. 

We emphasize that both the distributions of the level spacing Eqs.~\eqref{eq: Poisson Pr distribution} and ~\eqref{eq: GOE Pr distribution} 
require a large number of samples to be observed. This can be achieved by either having a large system and having access to its full energy spectrum (self-averaging property) or by having multiple disorders realizations. However, in experimental settings, it is often difficult, or nearly impossible, to access the full energy spectrum of a quantum system. 

Here, we want to analyze the effect of only resolving sub-manifolds of the energy spectrum. The dimension of a spectral $k$-th sub-manifold $\dim \mathcal{H}_k$, for a spin chain, is given by the decomposition
\begin{align}
    \dim \mathcal{H}&=2^N=\sum_{k=0}^N\mqty(N \\ k)=\sum_{k=0}^L\dim \mathcal{H}_k & \mqty(N \\ k)&=\frac{N!}{k!(N-k)!}.
\end{align}
In Figure~\ref{fig: matrix ensembles ideal}, we simulate the full spectrum for a matrix model of dimension $2^N$, with $N=8$. Subsequently, we then reduce to the lowest $d_\text{red}=37$ eigenvalues, which constitutes the third sub-manifold $d_\text{red}=1+8+28=37$. Notably, including the second manifold, we find that the spectral observables still host similar features for the matrix model (Fig.~\ref{fig: matrix ensembles reduced spectrum}A and B).

\subsection{Spectral form factor}

\begin{figure}[h!]
    \centering
    \includegraphics[width=\linewidth]{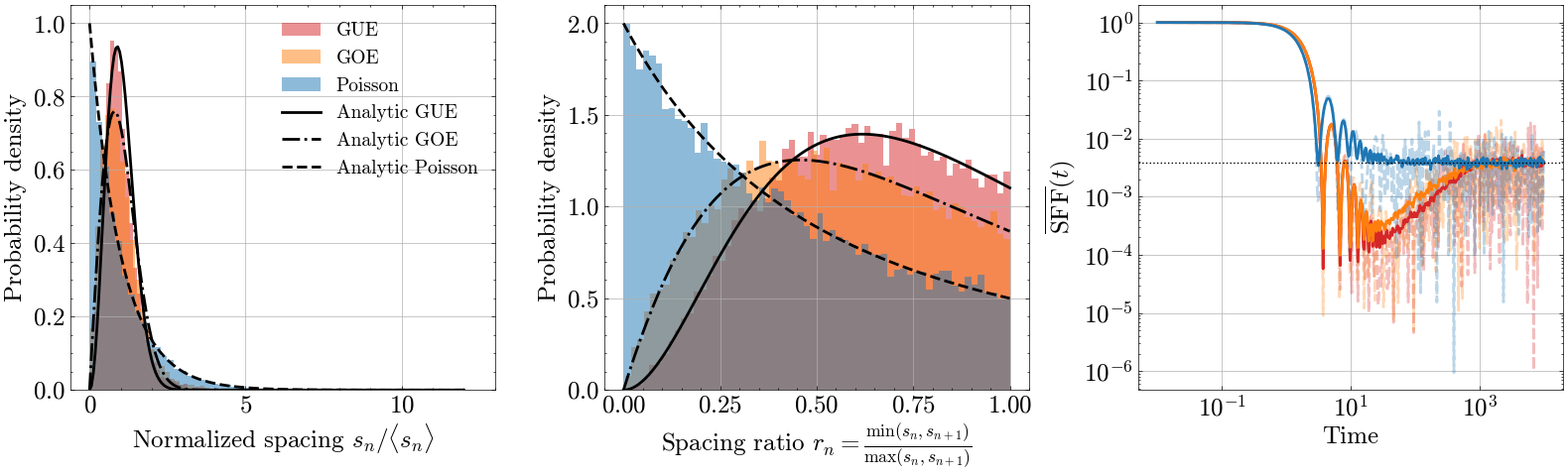}
    \caption{Different ways to capture eigenvalue statistics for a matrix of size $2^L\times 2^L$ with $L=8$: averaged spectral form factor (left), energy spacing distribution (middle), and energy spacing ratio distribution (right). In the averaged spectral form factor, we also plot a single instance in transparent font to illustrate the non-self-averaging of the SFF~\cite{Prange1997}. In orange, we show the statistics for the uniformly distributed eigenvalues, and in blue, the ones from a Gaussian unitary ensemble. Similarly, for the GOE, we plot the spectral measures in red. For all quantities, we draw the eigenvalues from their respective distribution and average over $N_\text{sim}=100$ samples. For the histograms, we choose a bin size of 60.}
    \label{fig: matrix ensembles ideal}
\end{figure}

Another observable with distinct features for integrable and chaotic systems is the spectral form factor $\text{SFF}(t)$, whose features we will discuss here.

The spectral form factor is defined as (using $\hbar=1$)~\cite{Altland2025}
\begin{align}
    \text{SFF}(t)\equiv \frac{1}{d^2}|\tr \hat{U}(t)|^2=\frac{1}{d^2}\sum_{n,m}e^{i(E_n-E_m)t}
\end{align}
for a Hilbert space of dimension $d=\dim \mathcal{H}$ with eigenvalues of the system Hamiltonian being $E_n$. Here we assume that the Hamiltonian is time-independent, which is the case in the adiabatic limit. Alongside the spectral form factor, we will separately introduce the average SFF as
\begin{align}
    \label{eqn supp: def SFF}
    \overline{\text{SFF}}(t)\equiv \frac{1}{d^2}\overline{\sum_{n,m}e^{i(E_n-E_m)t}}\approx \frac{1}{N_\text{sim}}\sum_{j=1}^{N_\text{sim}} \text{SFF}_j(t),
\end{align}
where the averaging $\overline{(\dots)}$ is an ensemble or disorder average. In the second line, we approximate the exact average over the eigenvalue distribution by the Monte Carlo approximation with $N_\text{sim}$ samples. 

Similar to the energy spacing ration, we can also observe the feature of eigenvalue repulsion, characteristic of chaotic systems, in the SFF (see Fig.~\ref{fig: matrix ensembles ideal}C). Similarly to the Poissonian case (blue curve), the GOE case shows an initial exponential decay and finalizes in the thermalized state, where the averaged SFF is given by $\overline{\text{SFF}}(t)\propto 1/d$. However, importantly, the decay of the GOE is stronger as it also shows a linear ramping for times around the \textit{Heisenberg time} $t_\text{H}=2\pi/\Delta$, where $\Delta=\min s_n$ is the minimal gap. The initial decay is related to the overall density of states $\rho(E)$, which for the Gaussian ensembles is known as the Wigner semicircle and for the Poissonian case is uniform. The linear ramp for the GOE is due to the eigenvalue repulsion.

Importantly, the SFF differs significantly from the spacing and spacing ratio distribution measures as it includes eigenvalue correlations beyond nearest neighbours, as can be seen in the fact that all combinations of eigenvalues are summed over in Eq.~\eqref{eqn supp: def SFF}. Therefore, the SFF captures global, long-range eigenvalue correlations as opposed to the local correlations in the eigenvalue spacing distributions. Additionally, the SFF is closely related to the survival probability~\cite{Das2025}
\begin{align}
    p_\text{S}(t)=|\langle \psi(0)|\psi(t)\rangle|^2=\sum_{n,m}|c_n|^2|c_m|^2e^{i(E_n-E_m)t},
\end{align}
where $c_n=\langle E_n|\psi(0)\rangle$ described the $n$-th
 component of the initial state $\ket{\psi(0)}$ written in the eigenbasis $\ket{E_n}$. We note that the return probability does equal the SFF if the initial state was prepared in a Gibbs state~\cite{MatsoukasRoubeas2024}. This provides the last interpretation of the SFF as the overlap between an initial state and its (possibly chaotic) time evolution.

\begin{figure}[h!]
    \centering
    \includegraphics[width=\linewidth]{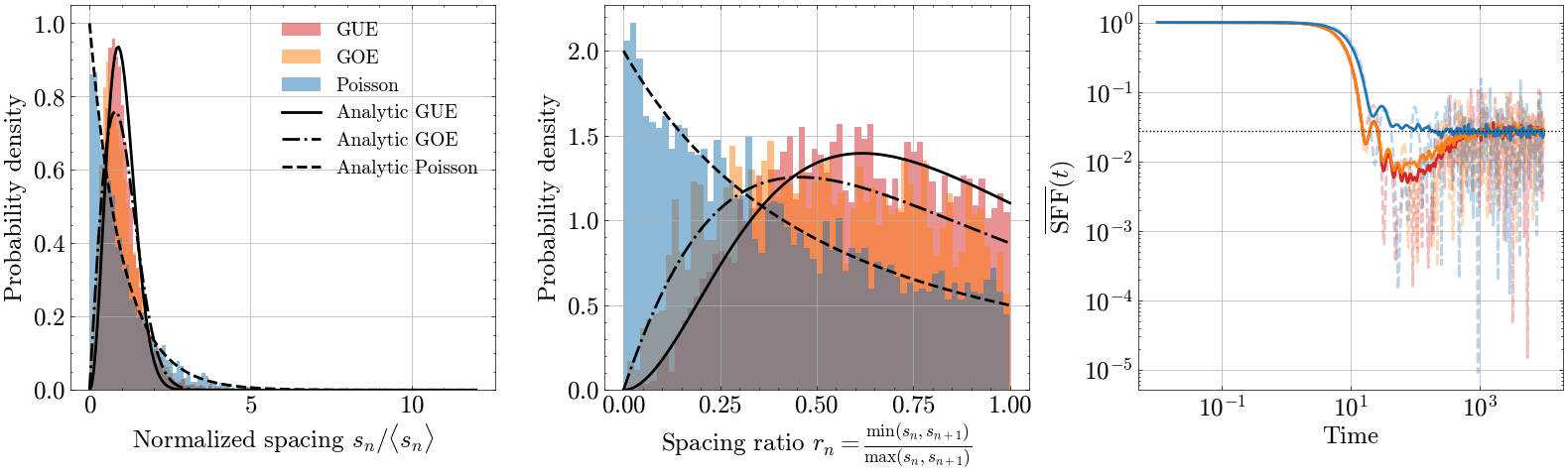}
    \caption{Eigenvalue statistics for a reduced spectrum: averaged spectral form factor (left), energy spacing distribution (middle), and energy spacing ratio distribution (right). We obtain the full spectrum and then, for the computation of the three observables, we reduce the spectrum to the second manifold $d_\text{red}=37<d=256=2^8$. For all quantities, we draw the eigenvalues from their respective distribution and average over $N_\text{sim}=100$ samples. For the histograms, we choose a bin size of 60.}
    \label{fig: matrix ensembles reduced spectrum}
\end{figure}

In the following, we elucidate further details about the early time behavior and the linear ramp, and compare numerical results to the observed experimental data.

\subsubsection{Early time behavior}
\begin{figure}
    \centering
    \includegraphics[width=0.6\linewidth]{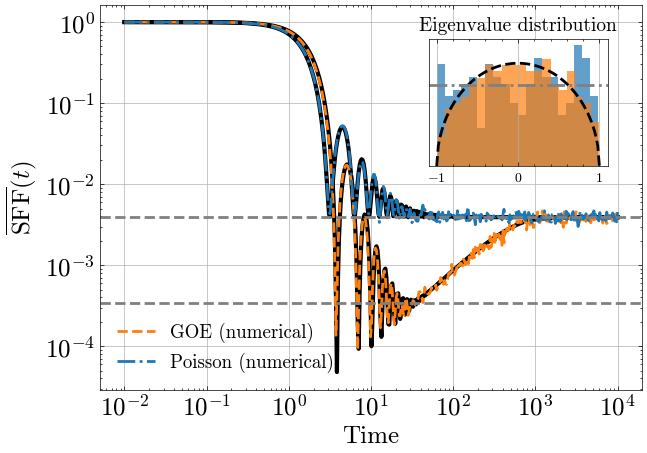}
    \caption{Analytics of the averaged SFF. In black, we illustrate the analytic expressions of the averaged SFF with GOE or uniform distribution of eigenvalues. Said (single-instance) distribution is shown in the inset. For the eigenvalue distribution we have chosen $E_0=1$ for both cases. For the chaotic case, we recover the Wigner semicircle law. The two gray dashed lines show the respective values at the dip, which are $\sim 1/d$ and $\sim 1/d^{3/2}$ for the uniform and chaotic case, respectively. The former agrees with both cases as their plateau constant.}
    \label{fig: analytics SFF}
\end{figure}

The early time behavior is characterized by an exponential decay, which only depends on the coarse-grained density of states $\rho(E)$ of each ensemble (more specifically, it depends on the Fourier transform $\Tilde{\rho}(t)$). For early times, one does not resolve fine-grained spectral information, hence there are no two-point eigenvalue correlations. Therefore,
\begin{align}
    \label{eq: average SFF formula}
    \overline{\text{SFF}}(t)&=\frac{1}{d^2}\overline{\sum_{n,m}e^{i(E_n-E_m)t}}
    =\frac{1}{d^2}\left(\overline{\sum_n 1}+\overline{\sum_{n\neq m}e^{i(E_n-E_m)t}}\right)\\
    &= \frac{1}{d^2}\left(d+\overline{\sum_{n\neq m}e^{iE_nt}}\;\overline{\sum_{n\neq m}e^{-iE_nt}}\right)\\
    &=\frac{1}{d^2}\left(d+d(d-1)\;\abs{\Tilde{\rho}(t)}^2\right),
\end{align}
where we used the fact that any higher-order eigenvalue correlations are negligible. This is valid at all times in the Poisson case, where there are no higher-order eigenvalue correlations. For uniformly distributed eigenvalues, we have $\rho(E)=1/2E_0 \text{ for } E\in [-E_0,E_0]$, and as such
\begin{align}
    \Tilde{\rho}(t)=\int_{-E_0}^{E_0}dE\, \left(\frac{1}{2E_0}\right)e^{-iEt}=\text{sinc}(E_0\,t).
\end{align}
Hence, we find that the SFF for the uniformly distributed eigenvalues takes the form~\cite{Prakash2021}
\begin{align}
    \overline{\text{SFF}}(t)\Big|_\text{Poisson}=\frac{1}{d}+\left(1-\frac{1}{d}\right)\text{sinc}^2(E_0\,t),
\end{align}
which holds for all times $t$ for the Poisson case. As shown in~\cite{Altland2025}, the initial decay part of the GOE can be derived from the Wigner semicircle, which takes the form
\begin{align}
    \rho_\text{GOE}(E)=\frac{2}{\pi E_0^2}\sqrt{E_0^2-E^2} \hspace{1cm}\text{for } E\in[-E_0,E_0].
\end{align}
We can compute the Fourier transform of the Wigner semicircle density of states
\begin{align}
    \Tilde{\rho}_\text{GOE}(t)=\int_{-E_0}^{E_0}dE\, \left(\frac{2}{\pi E_0^2}\sqrt{E_0^2-E^2}\right)e^{-iEt}=\frac{2 J_1(E_0t)}{E_0 t},
\end{align}
where $J_1(z)$ is the Bessel function of the first kind. This results in a decay proportional to $(J_1(E_0\,t)/(E_0 \,t))^2$ for times $t\ll t_\text{H}$. The full averaged SFF for the GOE is given by~\cite{Altland2025}
\begin{align}
     \overline{\text{SFF}}(t)\Big|_\text{GOE}=  \left(\frac{2J_1(E_0t)}{E_0t}\right)^2 - \frac{b_2(E_0t/\pi d)}{d} + \frac{1}{d},
\end{align}
where the two-point correlations are given by
\begin{align}
    b_2(t) =
\begin{cases}
1 - 2t + t \ln(1 + 2t) & \text{if } t \leq 1, \\
t \ln\left(\frac{2t + 1}{2t - 1}\right) - 1 & \text{if } t > 1.
\end{cases}
\end{align}
We can see in Figure~\ref{fig: analytics SFF} that both the numerical and analytical expressions (shown below as black curves) match for a system of size $d=2^8$. Note that initially (in the averaged case), one encounters resonance peaks in both cases, which are due to the zeros of the sine and Bessel functions, respectively. For a single instance of the eigenvalues, we find that the density of states reproduces the theoretical averaged limit of the Wigner semicircle well (See inset in Figure~\ref{fig: analytics SFF}), alluding to a good approximation of the density of states through a Monte Carlo approximation.  

\subsubsection{Eigenvalue repulsion from spectral form factor}
We can compute the averaged spectral form factor using the measure over the GUE
\begin{align}
    \overline{\text{SFF}}(t)&=\frac{1}{d^2}\int d\mu(H) \tr(e^{-iHt})\tr(e^{iHt}) & d\mu(H)&=dH \exp\left(-\frac{d}{2}\tr(H^2)\right),
\end{align}
where $d=2^L$. We can write the integral over the Hermitian matrix $H$ as an integral over eigenvalues by adjusting the measure 
\begin{align}
    dH=dU d\Lambda \,\Delta_\Lambda^2  \hspace{1cm} \text{ with }\Delta_\Lambda^2=\prod_{i<j}(\lambda_i-\lambda_j)^2,
\end{align}
where $U$ diagonalizes the matrix $H$, $\Lambda=(\lambda_1,\dots, \lambda_d)$ are the eigenvalues, and $\Delta_\Lambda^2$ is called the Vandermonde determinant. A non-zero Vandermonde determinant encodes level repulsion. The Vandermonde determinant arises due to the diagonalization $H=U\Lambda U^\dagger$, which we can derive from the Jacobian of the following infinitesimal expression $dH=d\Lambda + [dU, \Lambda]$, which we can evaluate in the eigenbasis $\{\ket{\psi_j}\}_{j=1}^d$ as
\begin{align}
    dH_{ij}=\delta_{ij}\,d\lambda_j+(\lambda_i-\lambda_j)\,dU_{ij} \hspace{1cm} \text{ where } X_{ij}=\mel{\psi_i}{X}{\psi_j} \hspace{0.2cm}\text{ for } X=H,U.
\end{align}
Taking the absolute value squared of the above quantity and noting that the cross-terms vanish, we find
\begin{align}
    \abs{dH_{ij}}^2=\delta_{ij}\,d\lambda_j^2+(\lambda_i-\lambda_j)^2\,dU_{ij}^2. 
\end{align}
The Jacobian is the determinant of the above quantity, which is
\begin{align}
    \det \left( \abs{dH_{ij}}^2 \right)=\prod_{i<j}(\lambda_i-\lambda_j)^2 \equiv  \Delta_\Lambda^2
\end{align}
This leads to an expression for the average SFF
\begin{align}
    \overline{\text{SFF}}(t)&=\frac{1}{d^2}\int dU d\Lambda \,\Delta_\Lambda^2 \exp\left(-\frac{d}{2}\sum_j \lambda_j^2-i\sum_{i,j}(\lambda_i-\lambda_j)t\right) \\
    &= \frac{\text{vol}[U(d)]}{d^2}\int d\Lambda \,\Delta_\Lambda^2 \exp\left(-\frac{d}{2}\sum_j \lambda_j^2-i\sum_{i,j}(\lambda_i-\lambda_j)t\right),
\end{align}
where we defined $\text{vol}[U(d)]=\int dU$ as the volume over the Haar measure. From this expression, we can see that the initial decay, i.e., for $t\to0$, is dominated by the distribution of single eigenvalues. For times $t\sim \mathcal{O}(t_\text{H})$, the oscillatory term allows for a finite integral over the Vandermonde determinant, which results in the known linear ramp~\cite{Altland2025, Liu2018, Winer2022}. One can derive the linear ramp through the sine-kernel method~\cite{Altland2025}. For simplicity, we focus on the connected part of the averaged SFF
\begin{align}
    \overline{\text{SFF}_c}\propto \int dEdE'\, R_2(E,E') e^{i(E-E')t},
\end{align}
where $R_2(E,E')=\overline{\rho(E)\rho(E')}-\delta(E-E')\rho(E)$ is the two-point spectral correlation function. In the large matrix size limit, the correlation function approaches the sine-kernel
\begin{align}
    R_2(E,E')\approx \left(\frac{\sin(\pi(E-E'))}{\pi(E-E')}\right)^2,
\end{align}
which crucially depends only on the energy difference originating from the Vandermonde determinant. As only the difference between eigenvalues matters, we may write~\cite{Altland2025}
\begin{align}
    \overline{\text{SFF}_c}\propto \int_{-\infty}^{\infty} ds\, \left(\frac{\sin(\pi s)}{\pi s}\right)^2 e^{ist}\propto t \hspace{1cm} \text{for }t<t_\text{H}
\end{align}
resulting in the linear ramp.

\subsubsection{Note on degeneracies for spectral analyses}
To analyze and conclude the correct spectral behavior (for both the gap ratio and the SFF), one needs to account for the underlying symmetries in the physical systems and subsequently work in a corresponding (irreducible) symmetry block. If one does not remove these symmetries, any spectral analysis will wrongly skew the gap ratio distribution and also SFF towards the Poissonian case, as one masks the eigenvalue repulsion in the case of Gaussian ensembles.

\subsection{Robustness analysis 8-spin chain SFF}

\begin{figure}
    \centering
    \includegraphics[width=0.9\linewidth]{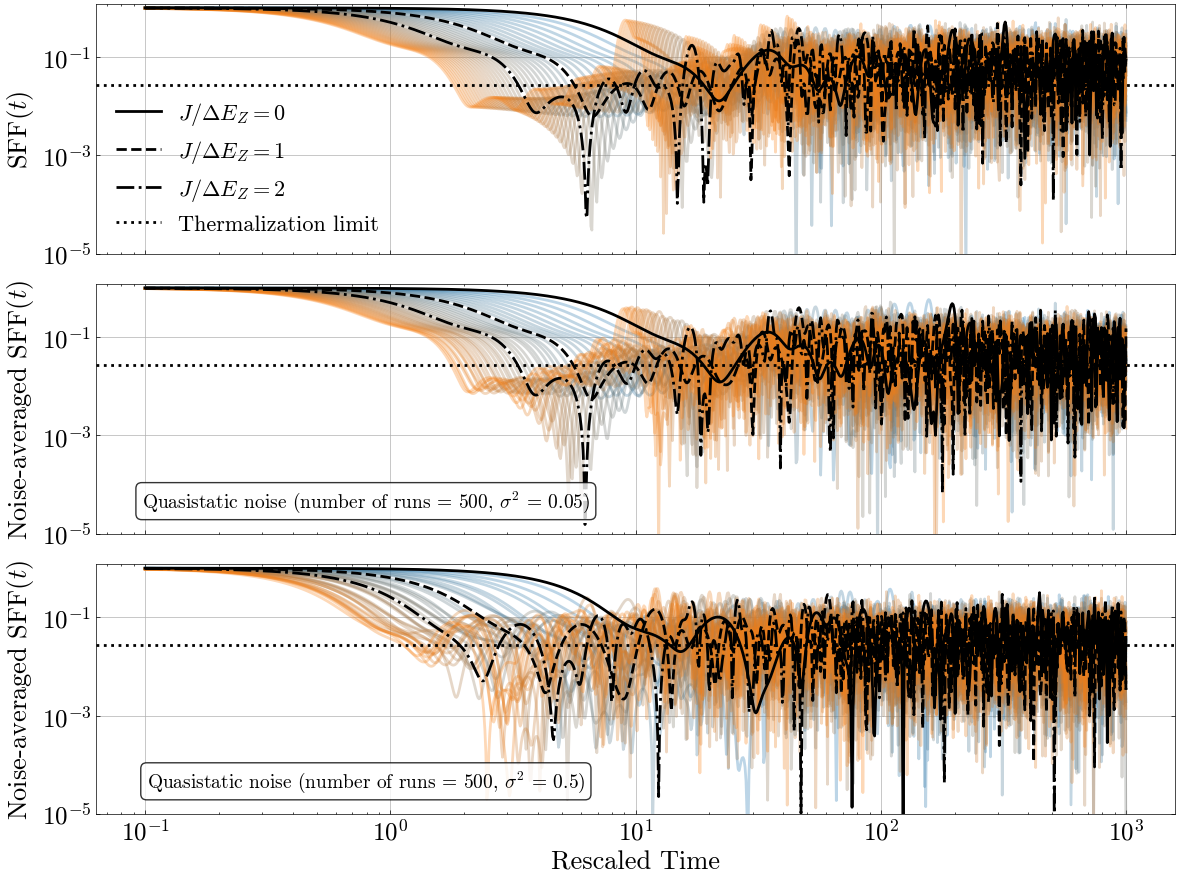}
    \caption{Spectral form factor from 8-spin chain with experimentally-extracted parameters without (top) and with (bottom) quasistatic noise. We add the noise to all parameters in the model (barrier voltage offset, g-factors, and spin-orbit tunnel coupling) via $\lambda \to \lambda(1+\epsilon)$, where we sample $\epsilon\sim \mathcal{N}(0,\sigma^2)$ and average over 500 realizations. To reduce exposure to outliers, we employ the median instead of the mean for the quasistatic noise analysis.}
    \label{fig: sff from experimental data}
\end{figure}

In this section, we aim to provide an argument for the statistical stability of the SFF under quasistatic noise. In particular, we aim to show the same trend for increasing exchange. The 8-spin chain hosts anisotropic Heisenberg exchange, where the exchange couplings are tuned through the barrier voltage
\begin{align}
    J_{ij}=J_{0,ij}\exp\Big(k_{ij}(b_{ij}-b_{0,ij})\Big),
\end{align}
Here, the $b_{ij}$ correspond to the virtualized barrier voltages, and $b_{0,ij}$ is an offset. In Figure~\ref{fig: sff from experimental data}, we compute the spectral form factor simulating the eigenvalue statistics for the experimentally-extracted Hamiltonian parameters. In addition to the individual single-instance spectral form factors for each exchange configuration, we also plot a quasistatic noise average. We take the median, instead of the mean, to circumvent outliers in the quasistatic noise analysis. Notably, for all three voltage conditions (with and without noise), we find that the spectral form factors recover roughly the features of the ideal spectral form factors in both regimes. Particularly, we find that the trend of a deeper dip for stronger exchange over disorder, at short times, is robust under quasistatic noise.

\clearpage

\section{Setup}
\label{sec:Setup}
All experiments are carried out in an Oxford Triton dilution refrigerator at a base temperature of 13~mK. The sample is mounted on a custom built printed circuit board (PCB) with a total of 100 DC, 28 microwave (up to $\SI{2}{\giga\hertz}$, and 2 RF-reflectometry lines. The microwave lines are connected to their corresponding DC lines via bias tees with a cutoff frequency of 10 kHz. Pulses to the gates are generated by Keysight M3202A arbitrary waveform geberators. The RF excitation for readout is generated by a custom build RF module, which is also employed to demodulate the reflected signal which is further digitized in the Keysight M3102A digitizer module. DC voltages are provided by a custom built serial peripheral interface (SPI) rack. The measurement software is based on the Python libraries Core-tools and Pulselib as well as custom written code. 

Each dot is defined by a plunger gate $p_i$, designed to control its electrochemical potential $\mu_i$, as well as barriers to adjacent dots $b_{ij}$, which control the exchange interaction $J_{ij}$. Plunger and barrier gate voltages are virtual throughout this work, meaning that $p_i$ is a linear combination of voltages on all gates calibrated to only control $\mu_i$. Similarly, the virtual barriers $b_{ij}$ exclusively control the interaction strength $J_{ij}$ (in our previous work, we have extensively discussed the virtualization process~\cite{Jirovec2025}). Except during initialization and readout, we operate at the symmetry point where $\mu_i = \mu_j$ for all dots. This ensures a first order protection against charge noise boosting the spin coherence time~\cite{Martins2016, Reed2016}. Nearest-neighbor exchange is induced by negative voltage pulses on the barrier gates at the symmetry point.

Initialization and readout are always carried out in two-spin subspaces. We, therefore, group the eight dots into four pairs: 1-2, 3-4, 5-6, and 7-8 to calibrate these steps. Details about the initialization process can be found in our earlier work~\cite{Farina2025} and in section~\ref{sec:Initialization_pulses}.

The readout scheme relies on Pauli spin blockade (PSB) which, for small external magnetic fields, can discriminate a singlet state $|S \rangle$ from all the triplet states $|T_- \rangle$,  $|T_0 \rangle$, and $|T_+\rangle$~\cite{Nurizzo2023,Kelly2025}.

\section{Initialization and readout pulses}
\label{sec:Initialization_pulses}
\begin{figure}
    \centering
    \includegraphics[width=0.95\textwidth]{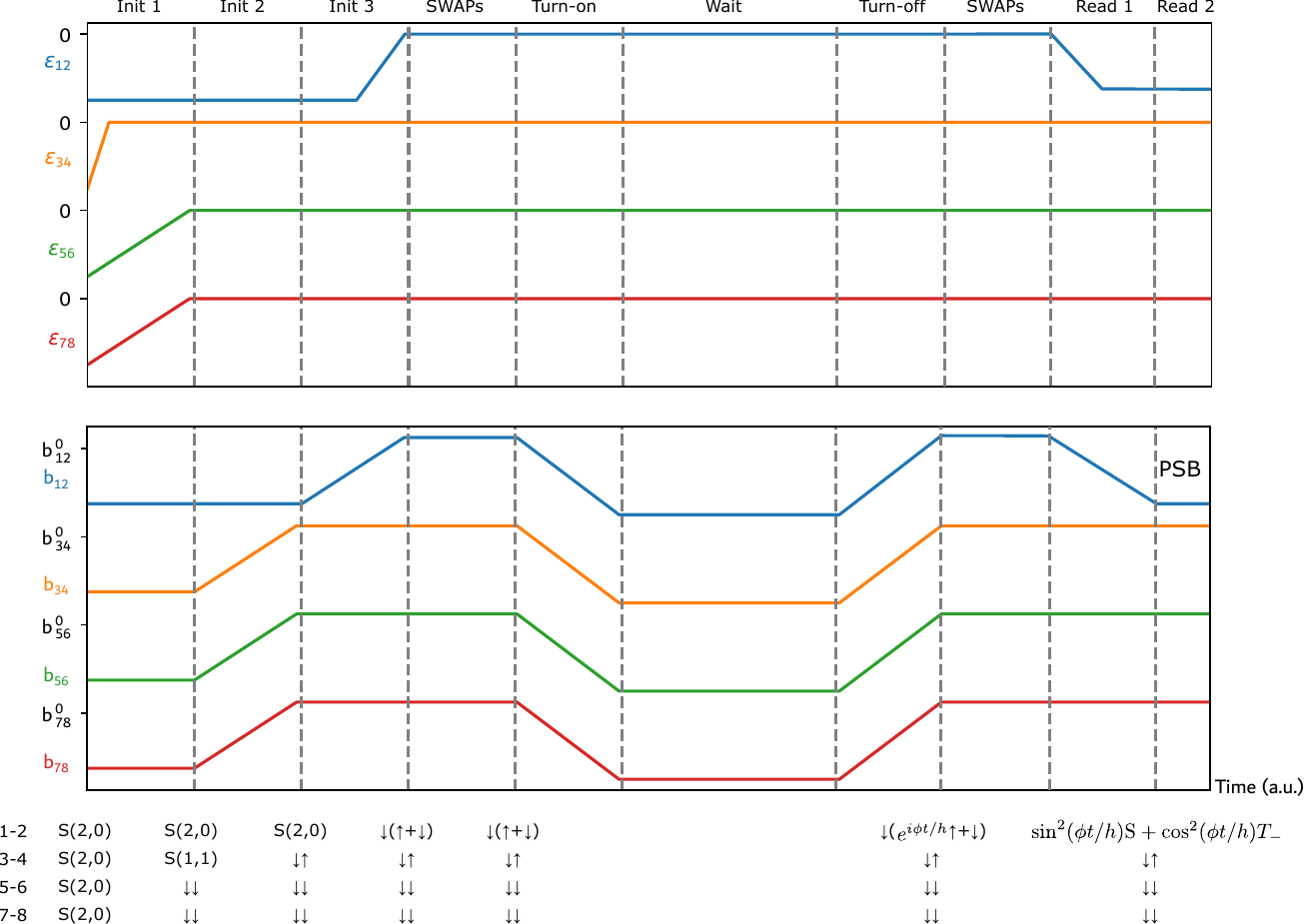}
    \caption{Typical pulses for an experiment involving all eight spins. The top panel depicts the pulses on detuning $\epsilon$, while the bottom panel depicts barrier pulses. Here, $b^0_{ij}$ is an offset voltage such that $J(b_{ij}=b^0_{ij}) = \SI{1}{\mega\hertz}$. For $b_{ij}>b^0_{ij}$ the exchange can be considered off. The graph is separated in phases explained in the text. After each phase and where possible, we further report the state in each pair.}
    \label{fig:experimental_pulses}
\end{figure}
Fig.~\ref{fig:experimental_pulses} depicts the pulses on detuning and barriers for a typical experiment involving all eight spins. In this particular case we initialize the state $\ket{\downarrow_1, \rightarrow_2, \downarrow_3, \uparrow_4, \downarrow_5, \downarrow_6, \downarrow_7, \downarrow_8}$. The graph is separated in parts which are explained in the following. After each part, marked by the gray dashed vertical lines, we also report the state in each pair (1-2, 3-4, 5-6, 7-8). At time 0, all detunings are large and the state in each pair is $S(2,0)$.

In the "Init 1" part, we quickly sweep $\epsilon_{34}$ to 0, while $\epsilon_{56}$ and $\epsilon_{78}$ are swept very slowly. After this part pair 3-4 is, therefore, in $S(1,1)$ (note that $b_{34}$ is very negative, inducing local exchange), while pairs 5-6 and 7-8 are both in $\downarrow\downarrow$.

In the "Init 2" part, we slowly close barriers $b_{34}$, $b_{56}$, and $b_{78}$, thereby closing the exchange and adiabatically transforming $S_{34}(1,1)$ into $\downarrow_3\uparrow_4$, while pairs 5-6 and 7-8 remain in their previous eigenstate ($\downarrow \downarrow$ is eigenstate for any exchange value).

In "Init 3" we semi-adiabatically ramp $\epsilon_{12}$ over the avoided crossing, while also closing the exchange with $b_{12}$. We now have prepared the desired state. 

During the "SWAPs" part, we can choose to apply sequential SWAP operations to move the $\uparrow$ or the superposition to a different location in the array. We note that in the figure no SWAP pulses are applied.

In the "Turn-on" part we slowly open all required exchanges by means of negative pulses on the barrier gates (additional pulses on $b_{15,23,48}$ are not shown). 

During the "Wait" part the phase accumulation takes place in the many-body sector.

In the "Turn-off" part we slowly close all required exchange interactions. After this, the phase information is stored in the superposition of spin 2, in this case.

If SWAPs were applied in the first "SWAPs" part, we now perform the required SWAPs to return the superposition state to the initial position.

Finally, we ramp the detuning semiadiabatically over the avoided crossing one more time to close the interference loop in "Read 1". The final state contains the desired information in the singlet probability which we readout in the PSB phase ("Read 2").

\clearpage

\section{SWAP calibration}
\label{sec:SWAP_calibration}
\begin{figure}[h!]
    \centering
    \includegraphics[width=0.95\linewidth]{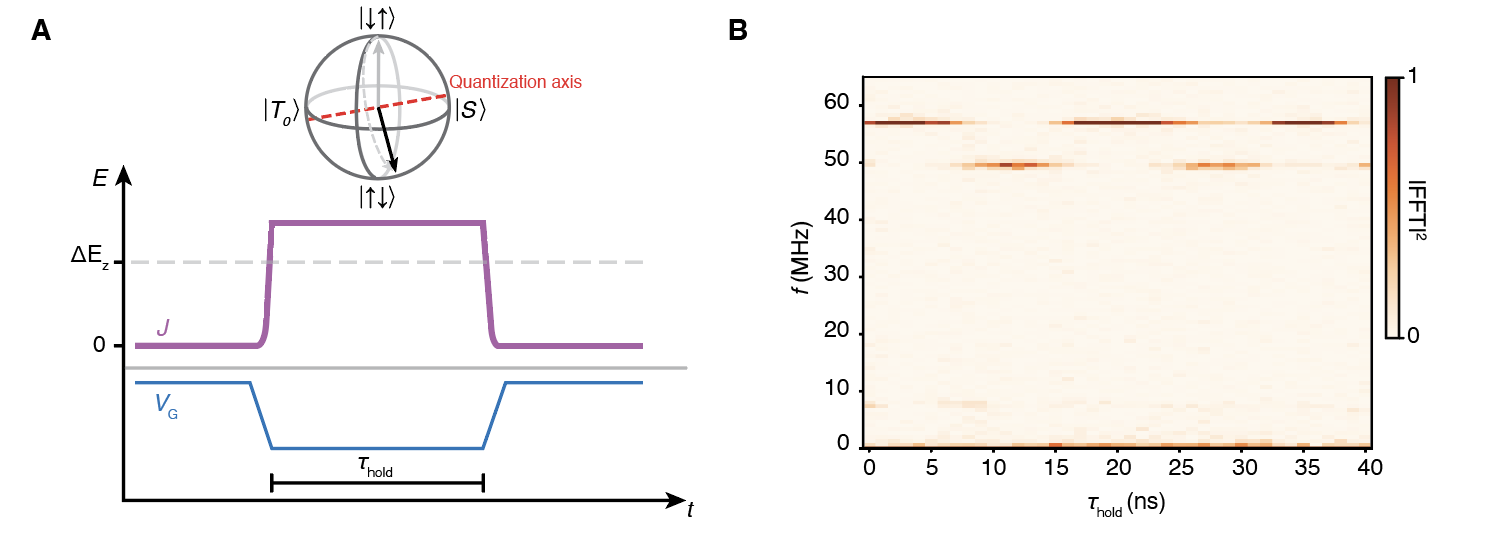}
    \caption{\textbf{A}, Schematic of the pulse sequence used used to implement SWAPS in Fig.~\ref{fig:BIMBS}F and expected state evolution in the $|\downarrow\uparrow \rangle$ - $|\uparrow\downarrow \rangle$ Bloch sphere. A voltage pulse $V_\text{G}$ on the barrier between two dots impose an exchange larger than the difference in Zeeeman energies $\Delta{}E_z$ between the two dots. \textbf{B}, FFT of the oscillations measured under the pulse scheme in A when calibrating the sequence for pair 1-2. Two clear frequencies are resolved, corresponding to the resonant frequencies of dot 1 (higher frequency) and dot 2 (lower frequency). An additional spurious frequency with low visibility is present around 8 MHz. This frequency corresponds to $\Delta{}E_z$ and is visible when the SWAP rotation is neither $~\pi$ nor $~2\pi$ and an effective superposition state is prepared.
    }
    \label{fig:SWAP1}
\end{figure}

As described in the main text, Landau-Zener passages selectively initialize the lowest g-factor spin in each dot pair in either spin down (slow ramp), a superposition state (fast ramp) or a spin up state (intermediate speed ramp. To prepare the higher g-factor spin in spin-up or superposition states, we implement SWAP operations through controlled exchange pulse sequences on barrier gates.
The data presented in Fig. \ref{fig:BIMBS}F were acquired using a single pulse as schematized in Fig. \ref{fig:SWAP1}A and described in previous work \cite{Farina2025}. Intuitively, this method can be understood considering a Bloch sphere in the $|\downarrow\uparrow \rangle$ - $|\uparrow\downarrow \rangle$ subspace. In the non-interacting regime used for state initialization, the quantization axis is along the $|\downarrow\uparrow \rangle$ and $|\uparrow\downarrow \rangle$ states. Increasing the exchange interaction among the two spins, the quantization axis tilts towards the $|S \rangle$ - $|T_0 \rangle$ states. Letting the system evolve in this regime leads to a rotation of the initial state in the Bloch sphere. To calibrate the exact time to obtain a half period rotation, we perform a Ramsey interferometry experiment varying the time the barrier is kept at the target voltage $\tau_{\text{hold}}$. The results of this calibration for the pair 1-2 is reported in Fig \ref{fig:SWAP1}B. An offset is present in the oscillations in the figure since the spins already start to swap during the fixed ramp time to the target gate voltage. For this reason at $\tau_{\text{hold}}$ = 0 a partial SWAP has already taken place and the dominant line corresponds to the resonance frequency of dot 1, higher than dot 2. This pulse scheme yields a relatively imprecise SWAP as can be seen by the residual oscillations around 8 MHz in Fig. \ref{fig:SWAP1}B and the spurious low frequency line in Fig. \ref{fig:BIMBS}F, both corresponding to $S-T_0$ oscillations. In fact, the imperfect SWAP prepares a superposition state $\alpha\ket{\downarrow\uparrow}+\beta\ket{\uparrow\downarrow}$, with $|\beta|^2\gg|\alpha|^2\neq0$. While acceptable for single operations where target frequencies remain dominant, fidelity degradation scales with sequential SWAPs, requiring more sophisticated pulse sequences for multi-SWAP state preparation protocols.

\begin{figure}[h!]
    \centering
    \includegraphics[width=0.95\linewidth]{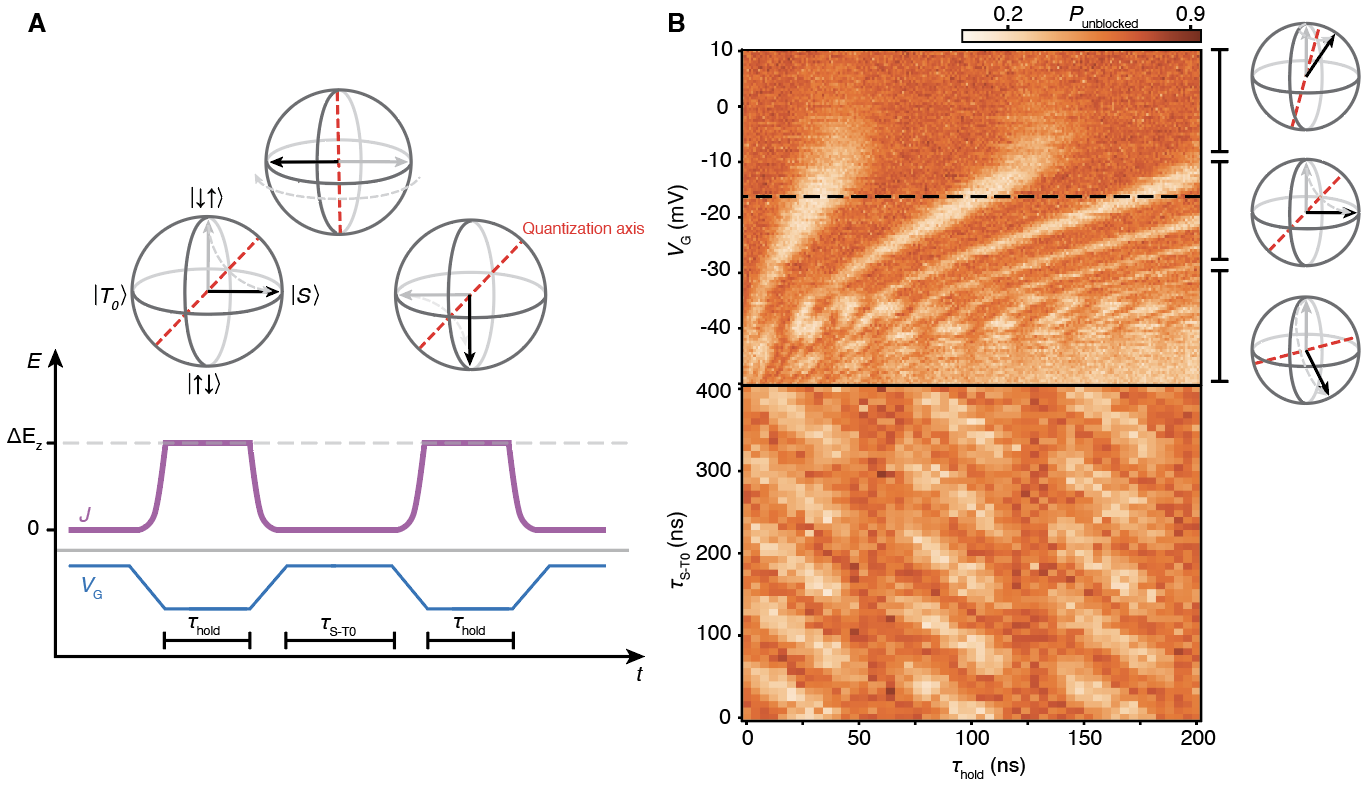}
    \caption{\textbf{A}, Schematic of the pulse sequence used for the SWAP operations and expected state evolution in the $|\downarrow\uparrow \rangle$ - $|\uparrow\downarrow \rangle$ Bloch sphere. Two voltage pulses $V_\text{G}$ on the barrier between two dots separated in time by $\tau_{S-T0}$ impose an exchange equal to the difference in Zeeeman energies $\Delta{}E_z$ between the two dots. \textbf{B}, Calibration of $V_\text{G}$, $\tau_{S-T0}$ and $\tau_{hold}$ for pair 5-6. From the first panel we select $V_\text{G}$ to be close to $\Delta{}E_z$ by choosing it to be in the region with large-amplitude oscillations with defined frequency. From the lower panel we calibrate $\tau_{S-T0}$ and $\tau_{hold}$ to correspond to a region with highly blocked PSB (brighter contrast).
    }
    \label{fig:SWAP2}
\end{figure}

To increase the SWAP fidelity we implemented a more sophisticated pulse sequence consisting of two single voltage pulses on the barrier gate separated by a waiting time $\tau_{S-T0}$ (Fig. \ref{fig:SWAP2}A) as realized in a previous work~\cite{Petit2022}. The sequence can be again understood considering the Bloch sphere in the $|\downarrow\uparrow \rangle$ - $|\uparrow\downarrow \rangle$ subspace. A first pulse impose an exchange $J$ between the two qubits equal to the difference in Zeeman energies $\Delta{}E_z$ between the two spins. This pulse tilts the quantization axis by 45 degrees. We can then calibrate $\tau_{\text{hold}}$ to perform a Hadamard gate. After this pulse the qubits are back to the non-interacting regime and held for a waiting time $\tau_{S-T0}$. During this time the quantization axis is along the $|\downarrow\uparrow \rangle$ and $|\uparrow\downarrow \rangle$ states. By calibrating $\tau_{S-T0}$, it is possible to obtain a Z-gate in the Bloch-sphere. Finally, a second pulse equal to the first one performs another Hadamard gate. Following the trajectory on the Bloch-sphere throughout these steps (see Fig.~\ref{fig:SWAP2}A), we see that this sequence takes $\ket{\uparrow\downarrow}$ to $\ket{\downarrow\uparrow}$ and viceversa. Additional verification confirms it implements a SWAP gate in the single spin basis.

Figure \ref{fig:SWAP2}B reports the calibration results for pair 5-6. In the top panel, we initialize the pair in the $|\downarrow\uparrow \rangle$ state and we acquire a 2D map by scanning the pulse heights $V_G$ and $\tau_{\text{hold}}$ with $\tau_{S-T0}$ = 0, and read out the $|\downarrow\uparrow \rangle$ probability \cite{Farina2025}. From this map it is possible to distinguish three different regimes as function of pulse height: at low voltages the visibility of oscillations is relatively small, at high voltages oscillations with multiple frequencies are present, resulting in a complex pattern, finally in the region between these two regimes it is possible to resolve oscillations with large contrast and a single frequency. The first regime corresponds to a small tilt of the quantization axis that barely influences the initial state. The large contrast oscillations are obtained when the quantization axis is close to 45 degrees, while larger pulses lead to two consecutive incomplete SWAPs resulting in a complex pattern. From this map we identify the target voltage for the pulses (dashed black line). Initializing and reading the state in the same way we then obtain a second map scanning $\tau_{\text{hold}}$ and $\tau_{S-T0}$ (lower panel in Fig. \ref{fig:SWAP2}B). The brightest areas in this map (fully blocked $|\uparrow\downarrow \rangle$ states) correspond to a well calibrated SWAP sequence and can be used to calibrate $\tau_{\text{hold}}$ and $\tau_{S-T0}$. All the SWAP operations used in the paper, except the one used in Fig. \ref{fig:BIMBS}F, were calibrated using this method.

\clearpage
\section{Larmor frequencies}
\label{sec:Larmor_freqs}
\begin{figure}[h!]
    \centering
    \includegraphics[width=0.95\linewidth]{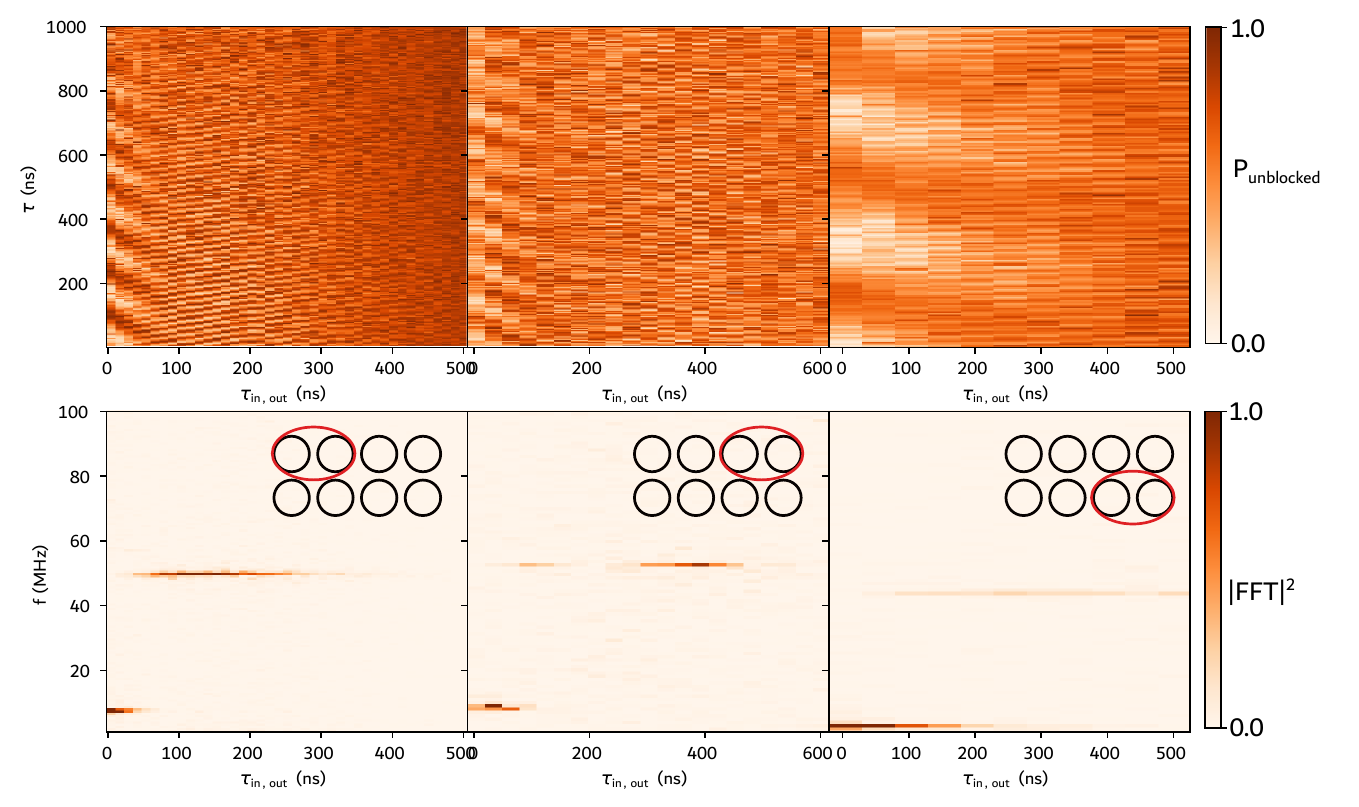}
    \caption{Determination of ramp times for pairs 12, 34, and 78. The top row plots the oscillations we measure as a function of $\tau_{ramp}$ for a pulse scheme as in Fig.~\ref{fig:BIMBS}A. The bottom row are the Fourier transforms of the top row where we can always observe a low frequency for short ramp times, corresponding to $S-T_0$ oscillations with a frequency $hf_{ij}^{ST_0} = \Delta g_{ij} \mu_B B$. At lon ramp times we can identify the Larmor frequency of the spin with the lower g-factor in each pair.
    }
    \label{fig:Init_ramps}
\end{figure}

We utilize the pulse schemes in Fig-~\ref{fig:BIMBS}A together with local SWAPs to map out the Larmor frequencies of the 8 spins in our system. We refer to the Larmor frequencies of the spins as the frequencies we measure when all $J$s are $\approx 0$. We record oscillations for up to 1~$\mu$s with a sampling rate of 500 MS/s, i.e. we record a point every 2 ns. After Fourier transforming the data we fit a Gaussian to the data which will detect the frequency peak we are interested in. As a first calibration step we need to determine $\tau_{ramp}$ which returns $P_{LZ}\approx 0.5$ for pairs 12, 34, and 78 (pair 56 is shown in Fig.~\ref{fig:BIMBS}B and C). The oscillations as well as their corresponding FFTs are plotted in Fig.~\ref{fig:Init_ramps}.

\begin{figure}[!h]
    \centering
    \includegraphics[width=0.9\linewidth]{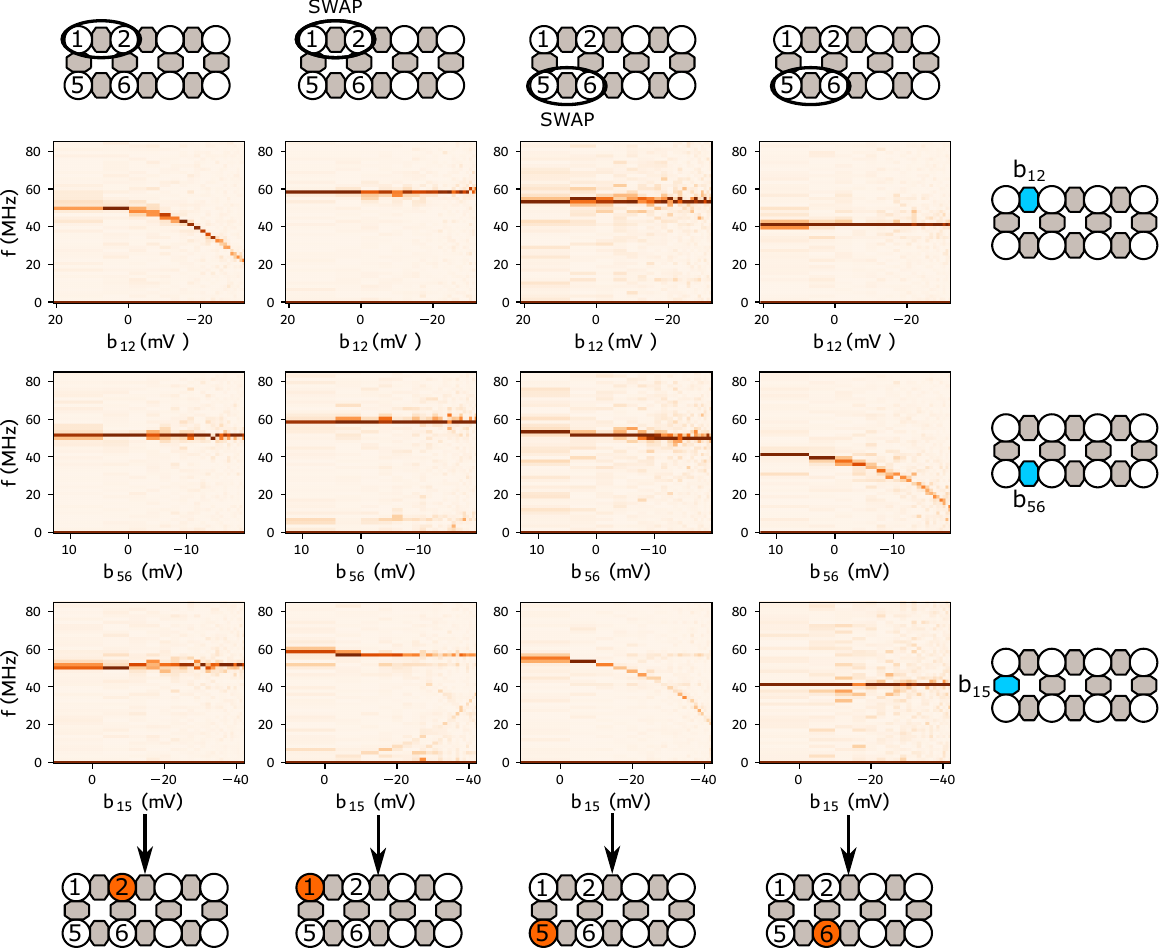}
    \caption{Mapping out the frequencies. For each column we prepare a certain spin in a superposition. The pair that is initialized is highlighted by an ellipse. When an additional SWAP is applied before and after the phase accumulation, we further mark the column with "SWAP". In each row, we scan a single barrier voltage highlighted by the sketches on the right side. From the dependencies of the frequencies on the barrier voltages we can infer the position of the spin in the array as explained in the text and highlighted by the sketches at the bottom of the figure.}
    \label{fig:Map_freqs}
\end{figure}

Subsequently, we perform local SWAPs as described in more detail in section~\ref{sec:SWAP_calibration}, to extract all the remaining Larmor frequencies. Lastly, the measured frequencies need to be mapped to the corresponding dots. To do this, we record the frequency of each dot as a function of a single barrier voltage. An example of such a procedure for the spin chain 2-1-5-6 can be seen in Fig.~\ref{fig:Map_freqs}. The top sketch in each column refers to the initialization and readout steps. The pair that is initialized is highlighted by an ellipse. Furthermore, if a SWAP before and after the phase accumulation is performed, an additional label 'SWAP' is present. On the right side we sketch which barrier is scanned in each row. For each pair of spins $i,j$ separated by barrier $b_{ij}$ only the lower resonance frequency will be strongly influenced by the exchange interaction. Of the four spins measured in the four columns of Fig.~\ref{fig:Map_freqs}, the spin shown in the fourth column has the lowest bare resonance frequency (with exchange switched off, i.e. positive barrier voltages). Therefore, its frequency should be altered by any adjacent exchange interaction. As we only observe a strong trend as a function of $b_{56}$ (second row) we can conclude that $b_{15}$ is not adjacent to this spin and, therefore, the spin we initialized in a superposition must be in dot 6. The other spin in the pair must be in dot 5 and we confirm this by observing a dependence of the frequency on $b_{15}$ (third row, third column). In pair $12$, one of the spins has a slightly lower resonance frequency than spin 5. However, it does not depend on $b_{15}$, but only on $b_{12}$ (first row, first column). We, therefore, conclude that this spin must reside in dot 2. The remaining spin is in dot 1 and does not react strongly to any barrier voltage as it has highest frequency.  

\begin{figure}[h!]
    \centering
    \includegraphics[width=0.9\linewidth]{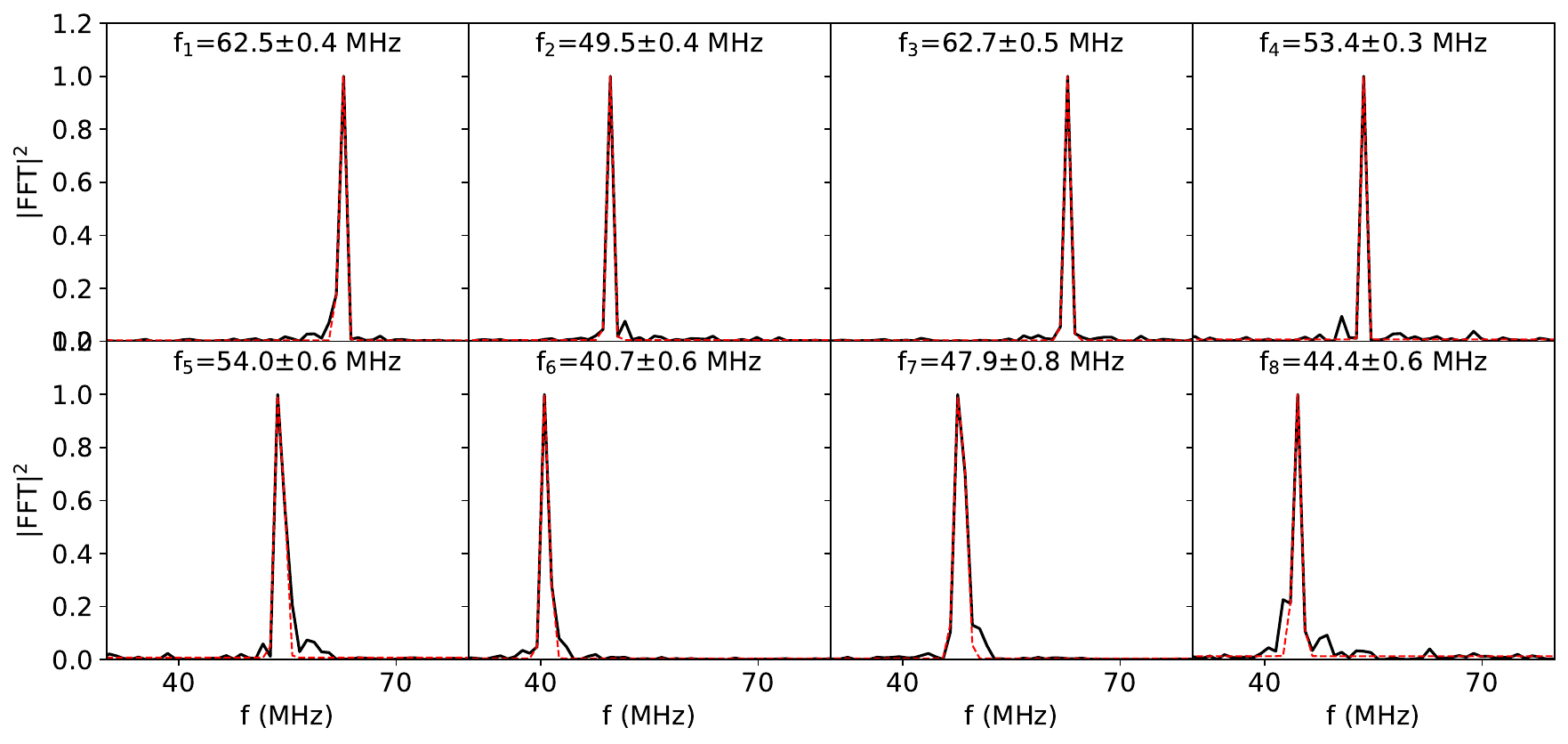}
    \caption{Larmor frequencies extracted for all eight spins. Spins 2, 4, 6, and 8 do not require SWAPs, while spins 1,3,5, and 7 do require SWAPs. The error is the full width at half maximum of the fits reported as dashed red lines.}
    \label{fig:Larmor_freqs}
\end{figure}

We utilize this information to reconstruct the disorder landscape of the device which we plot in Fig.~\ref{fig:Larmor_freqs}.

\clearpage
\section{Exchange and $\Delta_{SO}$ extraction}
\label{sec:Exchange_extraction}
In order to extract the relevant experimental parameters to simulate the eight-spin system, we perform interferometry experiments in two-spin subspaces. In this situation, the energy levels are well known and extensively studied for singlet-triplet qubits. For each pair of spins $ij$ we first measure the frequency of both spins as a function of $b_{ij}$. The minimum frequency we measure corresponds to $2\Delta_{SO}$ and occurs when $J(b_{ij})=\overline{E_Z}$. With this information we fit the full two-spin Hamiltonian to the data with $J$ as a free parameter and, for each barrier voltage, we extract the corresponding value of $J$.

\begin{figure}[h!]
    \centering
    \includegraphics[width=0.5\linewidth]{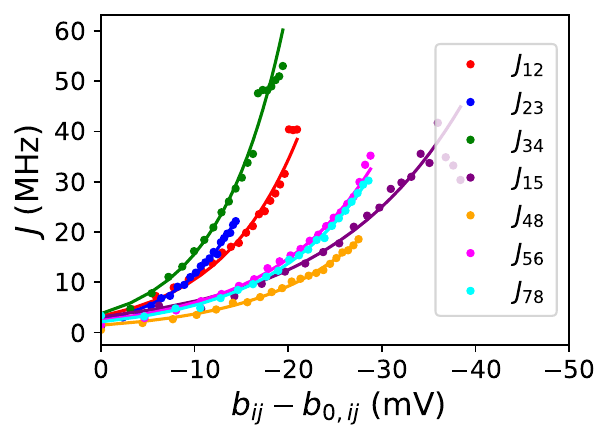}
    \caption{Exchange extracted from fits of eq.~\ref{eq:4spinHam} for two-spin subspaces as a function of the corresponding barrier voltage difference $b_{ij}-b_{0,ij}$, where $b_{ij}-b_{0,ij}=0$ refers to the barrier gate voltage where we acquired the data in Fig.~\ref{fig:Larmor_freqs}. The solid lines are fits to the empirical exchange formula given in the text.}
    \label{fig:All_Js}
\end{figure}

\begin{figure}[h!]
    \centering
    \includegraphics[width=0.8\linewidth]{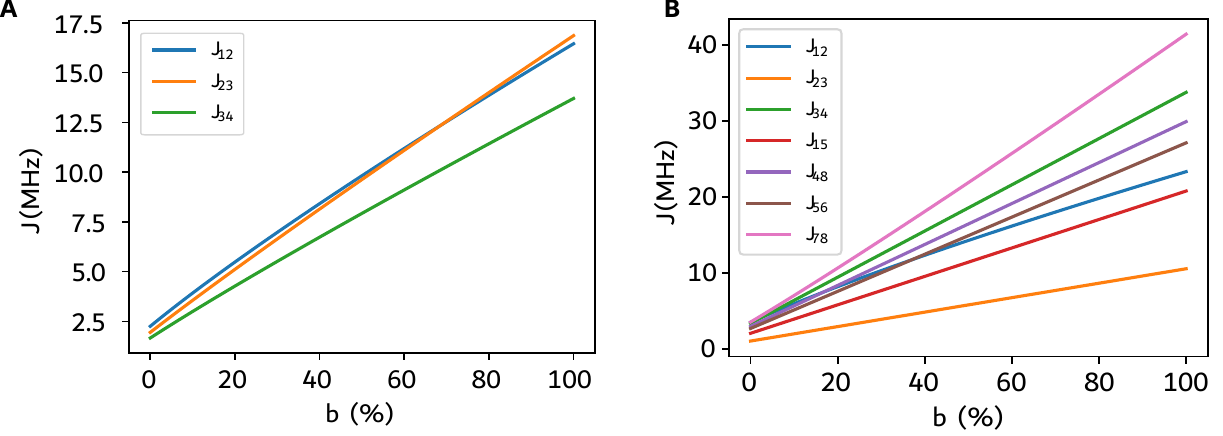}
    \caption{\textbf{Exchange values extracted from the four and eight spin chains in the main text}.
    (A) Exchange values for the chain 1-2-3-4 in Fig.~\ref{fig:4spin_spectroscopy} as a function of the applied barrier gate voltages $b$, where $b = 0\%$ corresponds to the case where $b_{ij}=b_{0,ij}$.
    (B) Exchange values for the eight spin chain in Fig.~\ref{fig:SFF} as a function of the applied barrier gate voltages $b$, where $b = 0\%$ corresponds to the case where $b_{ij}=b_{0,ij}$.}
    \label{fig:Exchange_fits_for_mainText_chains}
\end{figure}

We then fit the empirical formula $J(b_{ij})=J_0\exp{k(b_{ij}-b_{0,ij})}$ with $J_0= \SI{1}{\mega\hertz}$ and the two free parameters $k$ and $b_0$. The results are plotted in the Fig.~\ref{fig:All_Js}. We note that, in order to fit the four-spin and the eight-spin chain data in the main text, we slightly adjust the values of $b_{0,ij}$, which is a consequence of imperfect virtualization. We discussed the possible reasons in a previous work~\cite{Jirovec2025} and report the extracted exchange values in Fig.~\ref{fig:Exchange_fits_for_mainText_chains}.

Another parameter that can be slightly adjusted are the values for $\Delta_{SO}$. With regards to the theoretical model discussed in section~\ref{sec:Model}, we do not have acces to the angle $\gamma_{ij}$ which determines the degree of tilt in the principal g-tensor axis of two adjacent dots. In fact, we only probe the system in a single magnetic field direction and at a fixed magnetic field strength. Therefore, there is some level of uncertainty in the exact size of $\Delta_{SO}$. 

Finally, we generally assume that the effective g-factors are unchanged with gate voltage. However, for Ge spin qubits there is evidence that this is not always the case, making the fitting somewhat error-prone. In fact, it is likely that some of the deviations of the model from the multi-spin chain data can be explained by electric field induced g-tensor modulations.

\clearpage
\section{Additional spin chains}
\label{sec:Additional_spin_chains}
\begin{figure}[h!]
    \centering
    \includegraphics[width=0.85\linewidth]{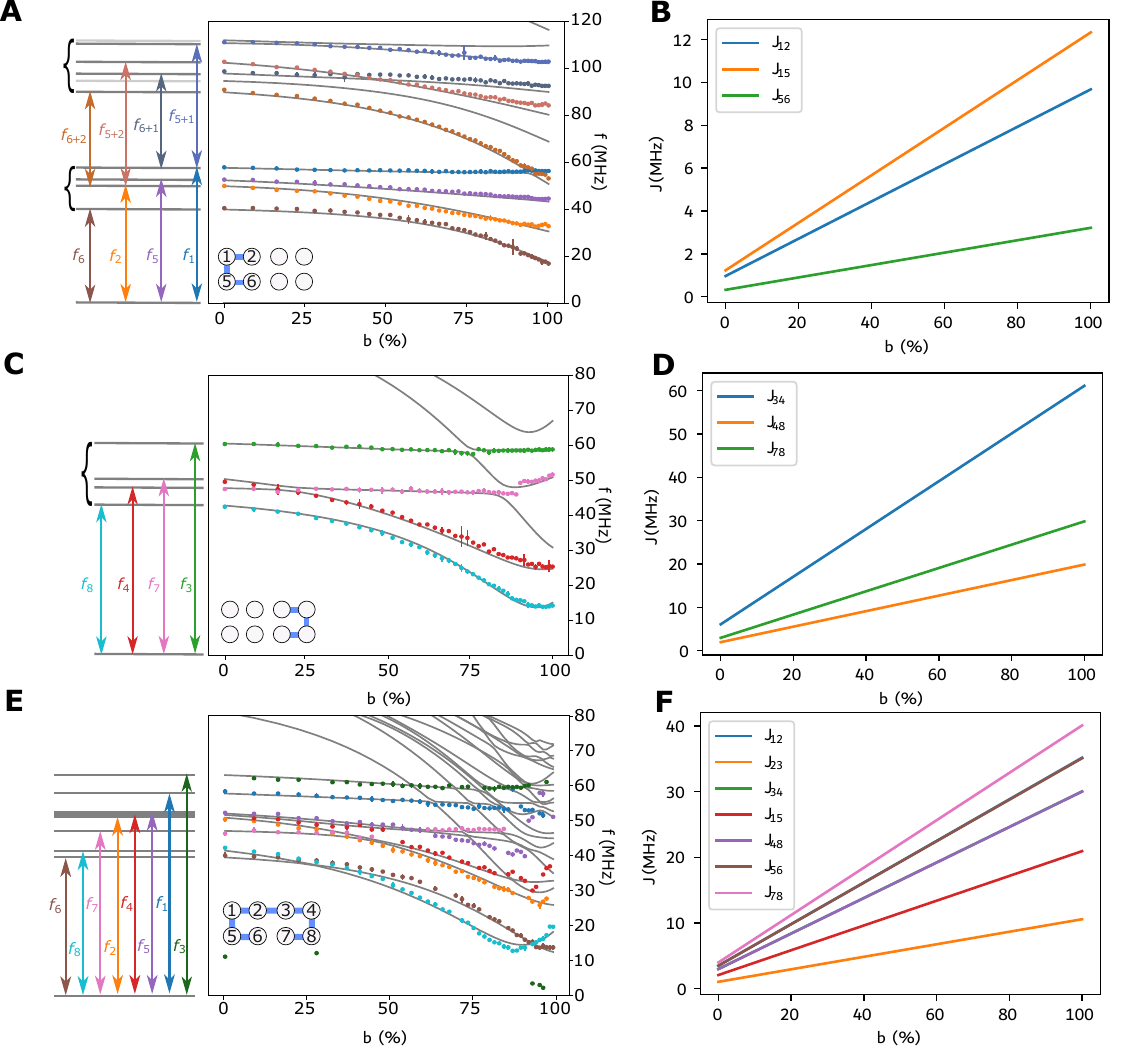}
    \caption{\textbf{Additional spin chains with model.}\textbf{A}. Chain formed by spins 2-1-5-6 where we also reconstruct most of the second spin manifold. The gray lines are the fitted model (eq.~\ref{eq:4spinHam} for four spins). \textbf{B}. Effective exchanges as extracted from the fit in A which shows a deviation from the target value.
    \textbf{C}. Chain formed by spins 3-4-8-7. The gray lines match the experimental data very well. Especially two avoided crossings, one for $f_3$ and one for $f_7$ are visible and can be explained by hybridization with higher spin manifolds. 
    \textbf{D}. Effective exchange extracted from the model for chain 3-4-8-7.
    \textbf{E}. Another instance of the full eight spin chain with the overlaid model which agrees very well with the data. 
    \textbf{F}. Effective exchange extracted from the model. We observe deviations from the target values for most exchanges. We utilize this plot to adjust the gate voltages and obtain a more homogeneous global exchange. }
    \label{fig:Other_chains}
\end{figure}

We report other measured spin chains in Fig.~\ref{fig:Other_chains}. Specifically, we show data for a chain formed by dots 2-1-5-6 (Fig.~\ref{fig:Other_chains}A), chain 3-4-8-7 (Fig.~\ref{fig:Other_chains}C), and another instance of the full eight-spin chain (Fig.~\ref{fig:Other_chains}E). The gray lines represent the modeled frequencies. Figures~\ref{fig:Other_chains}B, D, and F report the effective exchanges applied as extracted from the model and generally show some deviation from a homogeneous value we expect based on Fig.~\ref{fig:All_Js}. This can be attributed to residual barrier cross-talk. We utilize these measurements to adjust the gate voltages and calibrate a more homogeneous global exchange as a function of all $b_{ij}$ involved in the chains.

\clearpage
\section{Discussion on adiabatic ramps}
\label{sec:Adiabatic_ramps}
\begin{figure}[!h]
    \centering
    \includegraphics[width=0.95\textwidth]{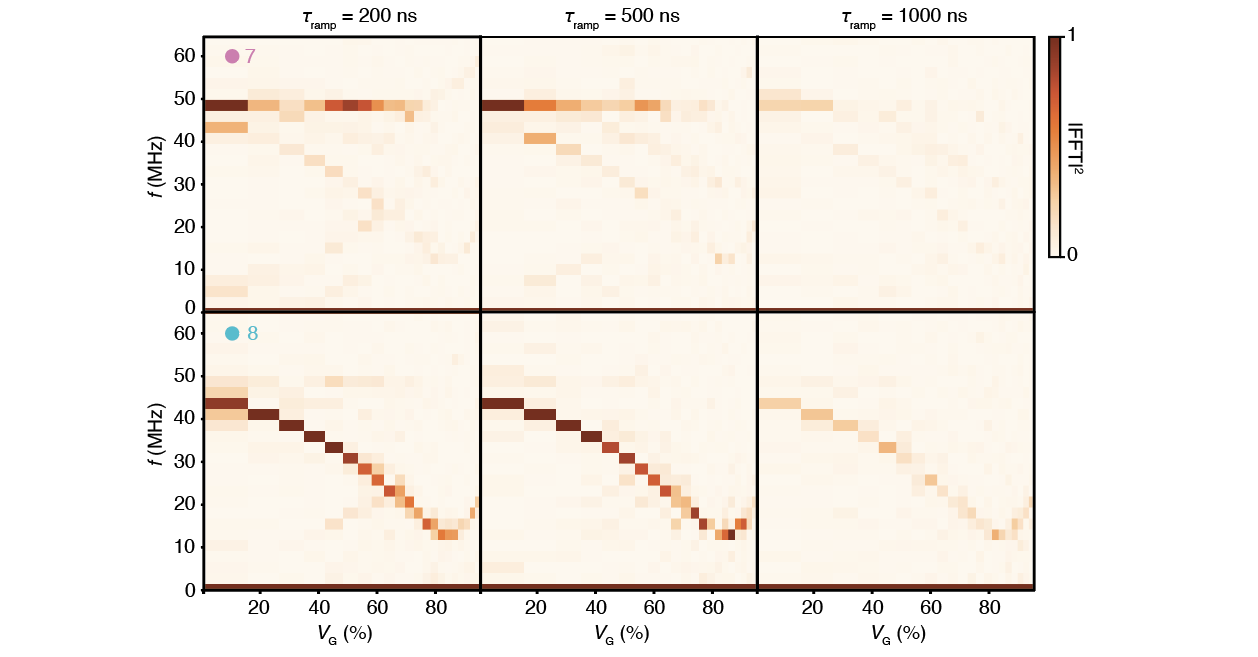}
    \caption{FFT of oscillations for the 8-spin chain reported in Fig. \ref{fig:SFF} measured at different ramp times $\tau_\text{ramp}$ initializing and reading-out dot 7 and 8 respectively. The data where normalized at the maximum signal among all the spectra in order to compare the FFT intensities.}
    \label{fig:adiabatic_exp}
\end{figure}

To ensure adiabatic ramps to the interacting regime, we acquired interferometric spectra as a function of ramp time $\tau_\text{ramp}$, reported in Fig. \ref{fig:adiabatic_exp} when preparing a single-spin superposition state in dot 7 or dot 8. For dot 8, at $\tau_\text{ramp}$ = 200 ns some spurious frequencies are present in the spectrum but the expected frequency is characterized by the largest FFT amplitude. Increasing $\tau_\text{ramp}$ to 500 ns results in a clearer spectrum, however at $\tau_\text{ramp}$ = 1000 ns the FFT intensity is largely reduced. We believe such behavior is due to decoherence effects, since we observe average $T_2^*\approx \SI{2}{\micro\second}$ for all the qubits at the symmetry point, and the total experimental time amounts to $>\SI{3}{\micro\second}$ (Initialization ramp, turn-on ramp, $\SI{1}{\micro\second}$ waiting time, turn-off ramp, and read-out ramp). For dot 7 we observe a similar trend; however, spurious frequencies are present in all three regimes. We attribute this to an imperfect initialization given by the small $\Delta{}E_z$ in the 7-8 pair, detrimental for the SWAP procedure~\cite{Seedhouse2021, Meinersen2024}. At $V_\text{G} >80\%$ the visibility of the FFT is drastically reduced, probably due to leakage to higher states. We observe similar behavior for the other spins in the chain (data not shown). For this reason, we limit the spectral form factor analysis to this value (see Fig. \ref{fig:SFF}). To further support our conclusions, we carry out extensive simulations as discussed in the following.

\begin{figure}[h!]
    \centering
    \includegraphics[width=0.95\textwidth]{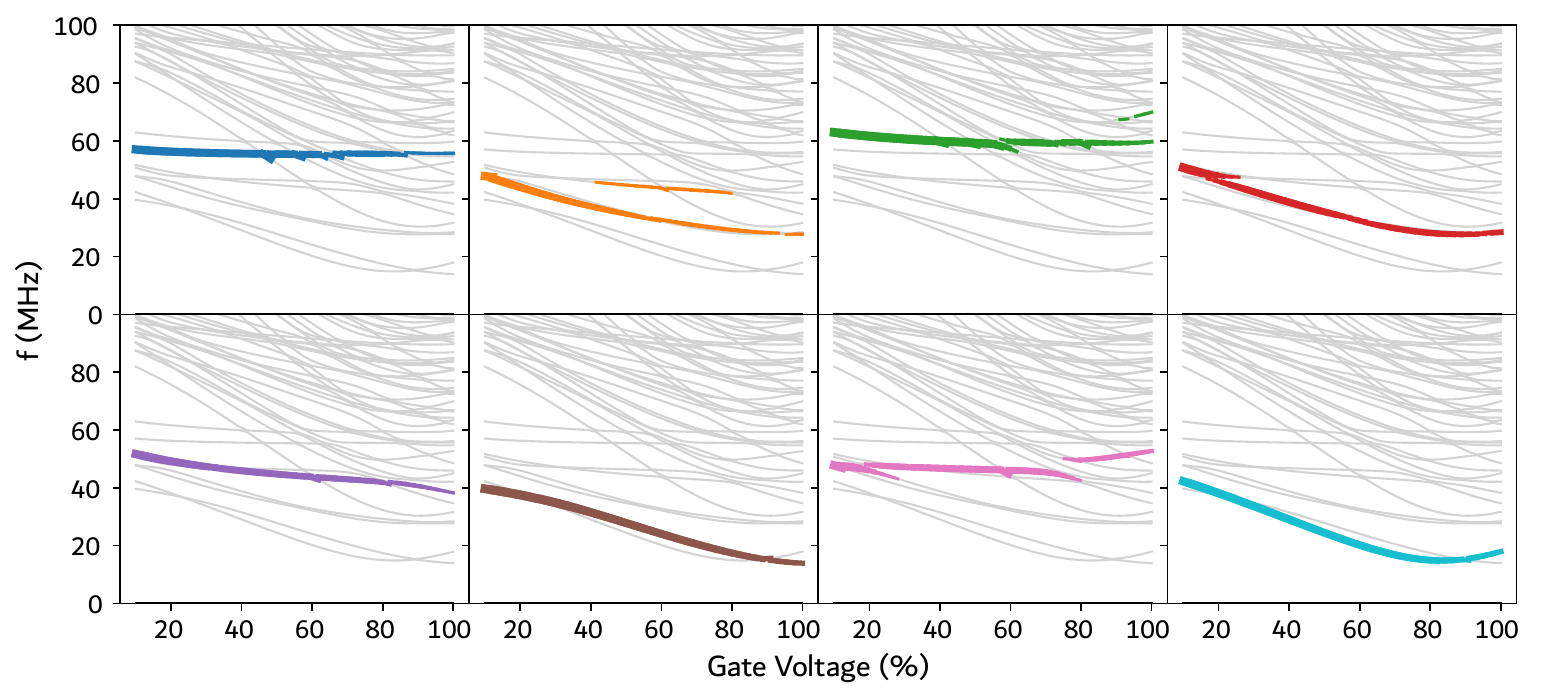}
    \caption{Simulated adiabatic eigenstates. The plots are geometrically orderer to reflect the position of the spin initialized in a superposition. The color code is the same as used in the main text to identify $f_i$ while the line thickness encodes the occupation probability. The x-axis represents the gate voltages that we applied also in the experiment in Fig.~\ref{fig:SFF}. }
    \label{fig:Simulated_adiabatic_eigenstates}
\end{figure}
The simulation tools at our disposal include adiabatic eigenstate deformation which we encode as a time dependent Hamiltonian in the python package Qutip. We write the time dependent Hamiltonian $H(t)= H_0+H_1\frac{t}{t_{max}}$.
Here $H_0 = \frac{1}{2}\sum_{i} (E_{Z,i} \sigma_{z,i} + \omega_{i,x} \sigma_{i,x})$ is the non-interacting part and $H_1 =  \frac{1}{4}  \sum_{\langle ij \rangle} J_{ij} ~\vec{\sigma}_i \cdot R_z (2 \gamma_{ij}) \vec{\sigma}_{j}$, and $t_{max}$ is the ramp-time.
The initial state is of the form $\ket{\downarrow_1\downarrow_2...\uparrow_i\downarrow_{i+1}..\downarrow_8}$ with only spin $i$ in an excited state. For each gate value, we then calculate the occupation probability of each eigenstate in the final Hamiltonian $H_1$ after the adiabatic turn on of exchange ($t_{max} = \SI{500}{\nano\second}$ in our case). To efficiently capture the exponential turn-on of exchange, we use a logarithmically spaced time axis in the simulations. The results are plotted in Fig. \ref{fig:Simulated_adiabatic_eigenstates} where each plot reflects the position of the excited spin in the device, and the color code follows the convention of the main text. The colored line thickness reflects the occupation probability, while the full spectrum is plotted in gray. For most cases, we indeed observe an adiabatic eigenstate deformation with no significant occupation of other states. For spin 2, 4, and 7 we do observe some leakage to other states which is also seen in the measurements. 

\clearpage
\section{Simulated spectral form factor}
\label{sec:Simulated_SFF}
\begin{figure}[h]
    \centering
    \includegraphics[width=0.9\linewidth]{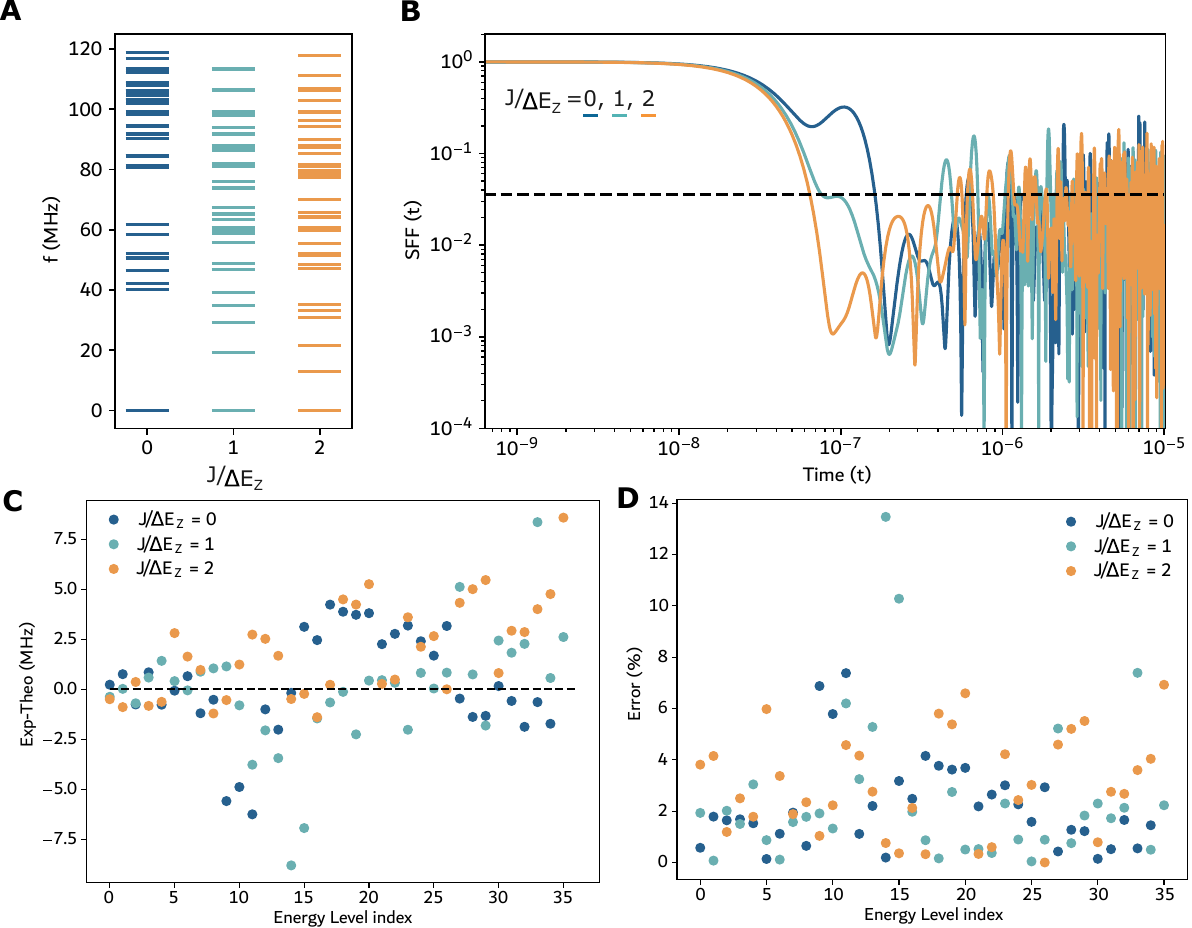}
    \caption{\textbf{A} Frequencies as extracted from model Hamiltonian with experimentally fitted parameters (Fig.~\ref{fig:SFF}B and Fig.~\ref{fig:Exchange_fits_for_mainText_chains}B) and simulated adiabatic ramps for the same exchange points as Fig.~\ref{fig:SFF}C. 
    \textbf{B} Theoretical spectral form factor for the levels extracted in \textbf{A}. While differences to Fig.~\ref{fig:SFF}D are noticeable, the early dip still develops as the interaction strength increases. 
    \textbf{C} Difference between the extracted (Fig.~\ref{fig:SFF}A) and the theoretically expected frequencies (panel \textbf{A}).
    \textbf{D} Relative error in percent of the absolute extracted frequency value. Except for two frequencies, we retain relative error rates below 10\% and, in most cases, even below 5\%. The remaining errors can be attributed to imperfect initialization, non-adiabatic ramps, and imperfect model parameters.
    }
    \label{fig:SFF_from_model_Ham}
\end{figure}

We utilize the adiabatic deformation also to calculate the expected frequencies in the eight-spin chain for higher manifolds with the model parameters used to fit the frequencies in Fig.~\ref{fig:SFF}B. This is necessary since any initial state will correspond to a state in the many-body sector, but we can not guarantee perfect adiabaticity due to the presence of small avoided crossings. The results are plotted in Fig.~\ref{fig:SFF_from_model_Ham}A, while Fig.~\ref{fig:SFF_from_model_Ham}B displays the corresponding SFF. Although we do see a difference between the experimentally extracted frequencies and the ones predicted by the model, we still observe the early dip developing for increasing interaction strength. In Fig.~\ref{fig:SFF_from_model_Ham}C we plot the difference between the measured frequency and the theoretically expected ones. Fig.~\ref{fig:SFF_from_model_Ham} shows the relative error in \% and we observe that for all but two frequencies we retain an error below 10\%, while in most cases the error is even below 5\%. These discrepancies can be explained by imperfect model parameters, not perfectly adiabatic ramps as well as imperfect initializations. While these errors are unavoidable as system sizes increase, importantly the SFF still returns information about the energy level correlations showcasing its aptitude to extract information about the state of the whole system from limited information.

\end{document}